\documentclass[prd,nofootinbib]{revtex4}
\usepackage{epsfig,amsmath}
\newcommand{\bra}[1]{\langle #1|}
\newcommand{\ket}[1]{|#1\rangle}
\newcommand{\braket}[2]{\langle #1|#2\rangle}

\def\d{{\rm d}}

\def\Pro{{\sf P}}

\newcommand{\tr}{{\rm tr}}

\newcommand{\e}{{\rm e}}

\newcommand{\omegav}{\boldsymbol{\omega}}

\newcommand{\sigmav}{\boldsymbol{\sigma}}

\newcommand{\phiv}{\boldsymbol{\phi}}
\newcommand{\Piv}{\boldsymbol{\Pi}}

\newcommand{\p}{{\rm p}}
\newcommand{\x}{{\rm x}}

\newcommand{\omt}{\frac{\omega}{T}}
\newcommand{\chiomt}{\chi\!\left( \omt \right)}

\begin{document}

\title{The ideal relativistic rotating gas as a perfect fluid with spin} 

\author{F. Becattini}\affiliation{Universit\`a di 
 Firenze and INFN Sezione di Firenze, Florence, Italy} 
\author{L. Tinti}\affiliation{Universit\`a di Firenze, Florence, Italy}

\begin{abstract}
We show that the ideal relativistic spinning gas at complete thermodynamical 
equilibrium is a fluid with a non-vanishing spin density tensor $\sigma_{\mu \nu}$.
After having obtained the expression of the local spin-dependent phase space 
density $f({\bf x},{\bf p})_{\sigma \tau}$ in the Boltzmann approximation, we derive
the spin density tensor and show that it is proportional to the acceleration tensor 
$\Omega_{\mu \nu}$ constructed with the Frenet-Serret tetrad. We recover the 
proper generalization of the fundamental thermodynamical relation, involving 
an additional term $-(1/2) \Omega_{\mu \nu} \sigma^{\mu \nu}$. We also show that 
the spin density tensor has a non-vanishing projection onto the four-velocity 
field, i.e. $t^\mu=\sigma_{\mu \nu} u^\nu \ne 0$, in contrast to the common 
assumption $t^\mu = 0$, known as Frenkel condition, in the thus-far proposed 
theories of relativistic fluids with spin. We briefly address the viewpoint
of the accelerated observer and inertial spin effects.
\end{abstract}

\maketitle

\section{Introduction}
\label{intro}

The study of relativistic fluids with spin has a long history. Among the early major 
contributors to the subject are Mathisson \cite{mathisson} and Weyssenhoff \cite{weys}.
The latter formulated a model enforcing the condition of vanishing projection of 
the spin density tensor over the four-velocity field, namely:
\begin{equation}\label{frenkel}
  t_\mu = \sigma_{\mu \nu} u^\nu = 0
\end{equation}
also known as {\em Frenkel condition}. This auxiliary constraint onto the fluid 
motion did not seem justified by established physical principles and for a system of
pointlike relativistic massive particles in general does not apply \cite{hagedorn}.
This condition was in fact critically reexamined in a later work by Bohm and Vigier 
\cite{bohm} where they proposed to extend the theory to a more general motion 
allowing the four-vector $t \ne 0$, yet without obtaining a conclusive quantitative 
formulation. A major step in formulating a theory of relativistic fluid with spin was 
made in 1960 by F. Halbwachs \cite{halb} who worked out a Lagrangian variational 
approach, yet keeping the Frenkel condition as auxiliary equation of motion. 

The hydrodynamics of relativistic fluids with spin has since become the subject 
of specialized literature in General Relativity \cite{vari}, especially in the 
framework of Einstein-Cartan gravity theory \cite{ec}. Most of these studies 
assume the Frenkel condition (\ref{frenkel}), which implies a considerable 
simplification in the calculations. Since this condition, as has been mentioned, 
does not derive from any major physical principle, it would be interesting to test 
it in a simple situation.

Indeed, a system which allows to perform a theoretical independent test exists: 
the macroscopic ideal spinning gas of particles with spin at full thermodynamical 
equilibrium. The condition of full thermodynamical equilibrium implies that this 
study can be carried out within the framework of statistical mechanics, independently 
of any previous formulation of a hydrodynamical theory and freely from the danger of 
begging the question.

One of us studied in a previous work \cite{becapicc} the physics of the ideal 
rotating relativistic gas of particles with spin (in the Boltzmann statistics) 
at full thermodynamical equilibrium obtaining the expression of the polarization 
four-vector and momentum spectra. In this paper, we resume that study and derive
the local expression of all thermodynamical quantities for this system seen as a
fluid. Particularly, we will obtain the expression of the spin tensor and the spin 
density tensor and show that the latter enters in the general thermodynamical equation 
relating proper entropy density, proper energy density and pressure, coupled to
the acceleration tensor. The nature of this additional term, which has been 
already extensively used in the aforementioned works \cite{vari,ec}, is that of 
a quantum correction to thermodynamics owing to the presence 
of spin as a further degree of freedom of particles. We will provide the expression 
of the single-particle spin-dependent phase space density showing that it is a 
non-diagonal matrix in spin coordinates. We will finally demonstrate that even for 
this simple system (indeed the simplest among all possible relativistic fluids) 
the four-vector $t$ defined in (\ref{frenkel}) is non-vanishing, thus violating 
the Frenkel condition.

The paper is organized as follows: in Sect.~\ref{eqgas} we summarize and further
explain the main results obtained in the work \cite{becapicc}. In Sect.~\ref{fluid} 
we calculate the local thermodynamical densities of the rotating gas viewed as
a macroscopic fluid. In Sect.~\ref{angvel} we introduce and derive the angular 
velocity tensor for a rotating fluid and in Sect.~\ref{spin} the spin tensor
showing its proportionality to the acceleration tensor. In Sect.~\ref{accel}
we briefly take the viewpoint of the accelerated fluid cell in the rotating
gas. 

Throughout this work, we will confine our attention to a special relativistic
fluid, i.e. we will disregard gravitational effects and assume a flat Minkowski
spacetime.

\subsection*{Notation}

In this paper we adopt the natural units, with $\hbar=c=K=1$.\\
The Minkowskian metric tensor is ${\rm diag}(1,-1,-1,-1)$; for the Levi-Civita
symbol we use the convention $\epsilon^{0123}=1$.\\ 
We will use the relativistic notation with repeated greek indices assumed to 
be saturated. Contractions will also be denoted by a dot for both vectors and tensors;
the saturated index in a contraction between a vector and a rank 2 tensor
is specified by the position of the vector, either on the left or right. 
Double contraction of rank 2 tensors $A$ and $B$ will also be denoted by a colon,
i.e. $A:B$.\\
Space-time linear transformations (translations, rotations, boosts) and SL(2,C) 
transformations are written in serif font, e.g. ${\sf R}$, ${\sf L}$; 
$[p]$ will denote the Lorentz or SL(2,C) transformation taking the timelike
vector $m \hat t = (m,0,0,0)$ to four-momentum $p$. Operators in Hilbert space will
be denoted by an upper hat, e.g. $\widehat {\sf R}$. Unit vectors will be denoted 
with a smaller hat e.g. $\hat {\bf p}$; even though the notation is unambiguous, 
we will make explicit mention of either possibility whenever confusion may arise. 

\section{Rotating relativistic gas at equilibrium}
\label{eqgas}

A macroscopic system with given angular momentum at complete thermodynamical 
equilibrium can only have a rigid velocity field:
\begin{equation}\label{rigid}
 {\bf v} = \omegav \times {\bf x}
\end{equation}
where $\omegav$ is a constant vector, as seen from the inertial system where
the system is at rest. This is a simple consequence of enforcing statistical 
equilibrium and is a well known result for non-relativistic systems \cite{landau} 
which can be easily extended to relativistic ones with the additional bound
$\| \omegav \times {\bf x} \| < 1$ (see Appendix B in 
ref.~\cite{becapicc}). Obviously, for an isolated system, $\omegav$ can be constant 
only if it is directed along one of the system's principal axes. Hence, 
thermodynamical equilibrium of a rotating body requires $\omegav$ to be parallel 
to ${\bf J}$ and, therefore, that the system's shape is symmetric for rotation
around $\omegav$.

The whole physics of an equilibrated system with finite angular momentum ${\bf J}$, 
chemical potential $\mu$ and temperature $T$ is fully contained in the density 
operator $\widehat \rho_{\bf J}$:
\begin{equation}\label{denmat1}
 \widehat\rho_{\bf J} = \frac{1}{Z_J} \exp [ (-\widehat H + \mu \widehat Q)/T] 
  \Pro_{\bf J} \Pro_V 
\end{equation}
where $\Pro_{\bf J}$ is the quantum Wigner projector onto a given state $\ket
{J,M}$:
\begin{equation}\label{wigner}
 \Pro_{\bf J} = \frac{1}{M(G)} \int_{SU(2)} \d g \; D^J(g^{-1})^M_M \; 
 {\widehat g}
\end{equation}
$Z_J$ is the partition function:
\begin{equation}\label{zj}
  Z_J = \tr \exp [ (-\widehat H + \mu \widehat Q)/T] \Pro_{\bf J} \Pro_V 
\end{equation}
and $\Pro_V$ is the projector onto {\em localized states} $\ket{h_V}$
\begin{equation}
  \Pro_V = \sum_{h_V} \ket{h_V}\bra{h_V}
\end{equation}
which form a complet set of quantum states for the system in its finite region
$V$. In a first quantization scheme, these states are a complete set of 
functions which vanish outside the domain $V$. On the other hand, in a quantum field 
theory, there are some complications which have been discussed extensively in 
refs.~\cite{beca1,beca2}, yet they do not essentially modify the first-quantization 
scenario (see also Appendix A).

The density operator (\ref{denmat1}) defines a grand-canonical ensemble with fixed
angular momentum. If both $V$ and ${\bf J}$ are large, it can be shown \cite{beca2,
becapicc} that this ensemble is equivalent (at least for first moments of 
statistical distributions) to the {\em rotational grand-canonical ensemble} whose 
density operator reads:
\begin{equation}\label{denmat2}
 \widehat\rho_{\omega} = \frac{1}{Z_\omega} \exp [ (-\widehat H + \mu \widehat Q + 
  \omegav \cdot \widehat {\bf J})/T] \Pro_V
\end{equation}
being $\omegav$ the angular velocity vector and $Z_\omega$ the relevant partition
function:
\begin{equation}\label{zomega}
  Z_\omega = \tr \exp [ (-\widehat H + \mu \widehat Q + 
  \omegav \cdot \widehat {\bf J})/T] \Pro_V
\end{equation}
The transition from (\ref{denmat1}) to (\ref{denmat2}), or from (\ref{zj}) to
(\ref{zomega}) can be carried out through a large $J$ limit of the Wigner rotation
matrices in (\ref{wigner}) and a subsequent saddle-point expansion in the angular 
variables defining the SU(2) parameters in the integral (\ref{wigner}); for details 
see refs.~\cite{beca2,becapicc}. For studying macroscopic rotating systems, the 
ensemble (\ref{denmat2}) is thus the most natural choice. It can be shown that 
indeed, as long as $\omega/T \ll 1$, the partition function (\ref{zomega}) can be 
written as the product of grand-canonical partition functions of infinitesimal cells 
moving with a velocity (\ref{rigid}), according to the general wisdom that 
equilibrated macroscopic systems must be rigidly rotating.

It is worth showing this in some detail for an ideal relativistic gas of massive
particles with spin $S$ in the limit of Boltzmann statistics, i.e. considering 
particle as distinguishable objects. In this case, we can simply write the partition 
function (\ref{zomega}) as a product of single-particle partition functions and 
the calculation of (\ref{zomega}) in fact reduces to compute matrix elements of 
single-particle compatible operators $\widehat h,\widehat q, \widehat {\bf j}$, like 
this:
\begin{equation}\label{matel1}
  \bra{p,\tau} \exp[-(\widehat h + \mu \widehat q + \omegav 
  \cdot \widehat {\bf j})/T] \Pro_V \ket{p, \sigma} =
  \e^{-\varepsilon/T + \mu q/T} \bra{p,\tau} \exp[\omegav 
  \cdot \widehat {\bf j}/T] \Pro_V \ket{p, \sigma} 
\end{equation}
where $\sigma$ and $\tau$ label polarization states \footnote{For the spin quantum 
numbers $\sigma$ we follow the convention in \cite{moussa}, where they are defined 
as eigenvalues of the projection of the Pauli-Lubanski vector divided by the mass 
onto the third space-like four-vector orthogonal to the particle four-momentum. For 
more details see ref.~\cite{moussa} and \cite{beca2}.}. The above equality requires
that, as mentioned, $[\widehat h, \widehat{\bf j}]= [\widehat q, \widehat{\bf j}]
= [\widehat q, \widehat h] = 0$.

The matrix element on the right hand side of eq.~(\ref{matel1}) can be calculated
by means of an analytical prolongation from imaginary $\omega$ values. After 
replacing $\omegav/T$ with $-i \phiv$ we have a rotation ${\sf R}_{\hat \omegav}(\phi)$
around the axis $\omegav$ of an angle $\phi$ appearing in the eq.~(\ref{matel1}).
We can then expand the matrix element on the right hand side as:
\begin{equation}\label{matel2}
  \bra{p,\tau} {\sf R}_{\hat \omegav}(\phi) \Pro_V \ket{p, \sigma} 
   = \sum_{\sigma'} \int \d^3 \p' \; \bra{p,\tau} \widehat{\sf R}_{\hat \omegav}
   (\phi) \ket{p',\sigma'} \bra{p,\sigma'} \Pro_V \ket{p, \sigma}
\end{equation}
Note that this expansion would have been impossible for the operator $\exp[\omegav 
\cdot \widehat {\bf j}/T]$ itself because, for real $\omega$, all its matrix 
elements calculated with momentum eigenstates turn out to be infinite. In other 
words, the correct procedure to make the analytic continuation of the matrix element 
is, first, go to imaginary $\omega$ values and thereafter insert an identity resolution 
like in (\ref{matel2}). The matrix element of the representation of a rotation 
is well known; it involves a Dirac's delta and a Wigner rotation matrix:
\begin{equation}\label{rota}
\bra{p,\tau} \widehat{\sf R}_{\hat \omegav}(\phi)\ket{p',\sigma'}=
 \delta^3\left({\bf p} - {\sf R}_{\hat \omegav}(\phi)({\bf p}')\right)
 D^S([{\sf R}_{\hat \omegav}(\phi)(p')]^{-1} {\sf R}_{\hat \omegav}(\phi)
 [p'])_{\tau \sigma'}
\end{equation}
where $[p]$ is defined as a Lorentz transformation (in fact its SL(2,C) 
correspondant) taking the timelike vector $m\hat t = (m,0,0,0)$ to $p$ and $D^S$ 
denotes, as usual, the (2S+1)-dimensional SL(2,C) representation (what is
commonly labelled as the (S,0) in fact). Obtaining the 
matrix element of the projector $\Pro_V$ over momentum eigenstates is not so 
straightforward. The calculation can be done in a quantum field theoretical framework 
\cite{beca2}:
\begin{equation}\label{prov} 
  \bra{p',\sigma'} \Pro_V \ket{p,\sigma} = \frac{1}{2} \, 
  \sqrt{\varepsilon \over \varepsilon'} \, \int_V \d^3 \x \; \e^{i {\bf x} \cdot 
  ({\bf p}-{\bf p}')} \left( D^S([p']^{-1}[p])+D^S([p']^{\dagger}[p]^{\dagger-1}) 
  \right)_{\sigma'\sigma} \bra{0} \Pro_V \ket{0}
\end{equation}
where the last factor $\bra{0}\Pro_V \ket{0}$, the vacuum expectation value of the 
projector $\Pro_V$, is an immaterial factor \cite{beca1,beca2} and tends to 1 for 
large volumes $V$ as $\Pro_V \to {\sf I}$; we will henceforth neglect it. The 
formula (\ref{prov}) gives the correct infinite volume limit $\Pro_V \to {\sf I}$,
that is the momentum eigenstate normalization. For finite volumes, a striking
feature is the presence of non-trivial matrix elements in spin coordinates;
their particular form can be justified also by requiring $\Pro_V$ to commute 
with rotation operators (see Appendix A).
 
Plugging (\ref{rota}) and (\ref{prov}) into (\ref{matel2}), and neglecting $\Pro_V$ 
vacuum expectation value, one obtains:
\begin{equation}\label{matel3}
  \bra{p,\tau} \widehat{\sf R}_{\hat \omegav}(\phi) \Pro_V \ket{p, \sigma} =
  \int_V \d^3 \x \; \e^{i {\bf x} \cdot 
  ({\bf p}-{\sf R}_{\hat \omegav}(\phi)^{-1}({\bf p}))} \;\; 
  \frac{1}{2} \left( D^S([p]^{-1} {\sf R}_{\hat \omegav}(\phi)[p])+
  D^S([p]^{\dagger} {\sf R}_{\hat \omegav}(\phi) [p]^{\dagger-1}) 
  \right)_{\tau \sigma}
\end{equation}  
In order to get to this expression, advantage needs to be taken of the unitarity
of the Wigner rotation, i.e.:
\begin{equation}\label{unitary}
 D^S([{\sf R}_{\hat \omegav}(\phi)(p')]^{-1} {\sf R}_{\hat \omegav}(\phi)
 [p']) =  D^S([{\sf R}_{\hat \omegav}(\phi)(p')]^\dagger 
 {\sf R}_{\hat \omegav}(\phi) [p']^{\dagger -1})
\end{equation} 
and the unitarity of ${\sf R}$ itself as an SL(2,C) matrix.
The equation (\ref{matel3}) can now be analytically prolonged to imaginary angles
yielding the final expression for the matrix element in (\ref{matel1}):
\begin{equation}\label{matel4}
  \bra{p,\tau} \exp[\omegav \cdot \widehat {\bf j}/T] \Pro_V \ket{p, \sigma} =
  \int_V \d^3 \x \; \e^{i {\bf x} \cdot 
  ({\bf p}-{\sf R}_{\hat \omegav}(i \omega/T)^{-1}({\bf p}))} \;\; 
  \frac{1}{2} \left( D^S([p]^{-1} {\sf R}_{\hat \omegav}(i \omega/T)[p])+
  D^S([p]^{\dagger} {\sf R}_{\hat \omegav}(i \omega/T) [p]^{\dagger-1}) 
  \right)_{\tau \sigma}
\end{equation}
The matrix element (\ref{matel4}) is the key ingredient to calculate the equilibrium 
single-particle phase space distribution and thence, the thermodynamics of the 
rotating Boltzmann gas; it is then worth some comment. 

Firstly, it should be noted that if the system's shape is symmetric by rotation 
around the $\hat\omegav$ axis, as has been mentioned, the projector $\Pro_V$ commutes
with the corresponding generator in the Hilbert space:
\begin{equation}\label{rotinva}
 [\Pro_V,\widehat{\bf j} \cdot \hat\omegav] = 0
\end{equation}
Therefore, the operator on the left-hand side of eq.~(\ref{matel4}) is hermitian 
and so, the expectation value on the left hand side ought to be real. This is easily 
seen for the spin matrices on the right-hand-side as ${\sf R}_{\hat \omegav}
(i \omega/T) = \exp[\omegav \cdot{\bf J}/T]$ is hermitian. Showing that the integral 
on the right-hand-side of eq.~(\ref{matel4}) is real requires a more elaborate 
calculation which is carried out in Appendix B. 

Secondly, it should be pointed out that the matrix element in (\ref{matel1}) has 
a spacial integral form which allows us to readily write a phase-space distribution:
\begin{equation}\label{phsp}
  f({\bf x},{\bf p})_{\tau \sigma} = \lambda \, \e^{-\varepsilon/T}
  \e^{i {\bf x} \cdot  ({\bf p}-{\sf R}_{\hat \omegav}(i \omega/T)^{-1}({\bf p}))} \;\;   
  \frac{1}{2} \left( D^S([p]^{-1} {\sf R}_{\hat \omegav}(i \omega/T)[p])+
  D^S([p]^{\dagger} {\sf R}_{\hat \omegav}(i \omega/T) [p]^{\dagger-1}) 
  \right)_{\tau \sigma}
\end{equation}
where $\lambda = \exp[\mu q/T]$ is the fugacity. Unlike its integral over the system's
region, the function in (\ref{phsp}) has, strictly speaking, no straightforward
physical meaning and it does not need to be real or positive. Eventually, one could
could calculate from it some more meaningful function like the Wigner's function. 
However, for very large systems, one expects the function (\ref{phsp}) to become
the actual phase-space single-particle distribution. For a cylindrically symmetric
system to be macroscopic, the product $T R$, being $R$ its maximal transverse radius, 
is supposed to be large, namely:
$$
  T R \gg 1
$$  
Hence, since a relativistic bound $\omega R < 1$ is implied in eq.~(\ref{rigid}), for 
proper macroscopic systems (and indeed for most practical purposes) the ratio between 
$\omega$ and $T$ is small, i.e.:
\begin{equation}\label{approx}
 \frac{\hbar \omega}{KT} \ll 1
\end{equation}
where we have purposely restored the natural constants. Consequently, the difference
between momenta in the exponent of (\ref{phsp}) can be well approximated by the
lowest order term in $\omega/T$, i.e.:
\begin{equation}\label{vectdiff}
 {\bf p} - {\sf R}_{\hat \omegav}(i \omega/T)^{-1}({\bf p}) = {\bf p} - 
 \left[ \cosh \omt \, {\bf p} - i \sinh \omt \, \hat \omegav \times {\bf p} + 
 (1-\cosh \omt) \, {\bf p}\cdot \hat \omegav \hat \omegav \right] \simeq
 i \omt \hat \omegav \times {\bf p}
\end{equation}
so that the phase-space distribution function (\ref{phsp}) reduces to:
\begin{eqnarray}\label{phsp2}
  f({\bf x},{\bf p})_{\tau\sigma} &=& \lambda \, \exp \left[ -\varepsilon/T 
  - {\bf x} \cdot (\omega \times {\bf p})/T \right] 
  \frac{1}{2} \left( D^S([p]^{-1} {\sf R}_{\hat \omegav}(i \omega/T)[p])+
  D^S([p]^{\dagger} {\sf R}_{\hat \omegav}(i \omega/T) [p]^{\dagger-1}) 
  \right)_{\tau \sigma} \nonumber \\
  &=& \lambda \, \exp \left[ -\varepsilon/T + {\bf p} \cdot (\omega \times 
  {\bf x})/T \right] \frac{1}{2} \left( D^S([p]^{-1} {\sf R}_{\hat \omegav}
  (i \omega/T)[p])+ D^S([p]^{\dagger} {\sf R}_{\hat \omegav}(i \omega/T) 
  [p]^{\dagger-1}) \right)_{\tau \sigma}
  \nonumber \\
  &=& \lambda \, \exp \left[ -\varepsilon/T + {\bf p} \cdot {\bf v}/T \right] 
  \frac{1}{2} \left( D^S([p]^{-1} {\sf R}_{\hat \omegav}
  (i \omega/T)[p])+ D^S([p]^{\dagger} {\sf R}_{\hat \omegav}(i \omega/T) 
  [p]^{\dagger-1}) \right)_{\tau \sigma}
\end{eqnarray}
where we have used eq.~(\ref{rigid}). The eq.~(\ref{phsp2}) is the unnormalized
single-particle phase-space distribution for the ideal rotating relativistic Boltzmann 
gas, seen by the inertial frame, and it turns out to be a non-diagonal matrix in 
the spin coordinates. The eq.~(\ref{phsp2}) is one of the main results of 
this work and clearly shows that for a rotating gas the polarization states are 
not evenly populated. Furthermore, if the total number of particles is not fixed, 
(\ref{phsp2}) coincides with the density of particles in phase-space, including 
spin degrees of freedom. In order to obtain the properly called phase-space density 
in $({\bf x},{\bf p})$ we have to take the trace of the matrix (\ref{phsp2}):
\begin{equation}\label{phsp1}
  f({\bf x},{\bf p}) = \sum_\sigma f({\bf x},{\bf p})_{\sigma \sigma}
  = \lambda \,  \e^{-\varepsilon/T + {\bf p} \cdot {\bf v}/T} \chiomt
\end{equation}  
being:
\begin{equation}\label{chiomegat}
\chiomt \equiv \tr D^S({\sf R}_{\hat \omegav}(i \omega/T)) =
 \frac{\sinh(S+\frac{1}{2})\omt}{\sinh(\frac{\omega}{2T})}  
\end{equation}
The exponent in (\ref{phsp1}) can be rewritten in a covariant form by introducing 
the temperature four-vector:
\begin{equation}\label{temp}
 \beta = \frac{1}{\gamma T} (\gamma, \gamma {\bf v}) = \frac{1}{\gamma T} u
  = \frac{1}{T_0} u
\end{equation}
being $\gamma= (1 - v^2)^{-1/2}$. Thereby, the invariant local temperature 
$T_0 = 1/\sqrt{\beta^2}$ measured in the comoving frame differs by a factor $\gamma$ 
from the global temperature measured by the inertial observer. Since $\gamma$ 
increases as a function of the radial distance, in a rotating gas at equilibrium
there will be a steady gradient of local temperature; since also $\mu/T$ must be
constant, there will also be a steady gradient of the local chemical potential,
i.e. the local chemical potential will be given by 
\begin{equation}\label{mu}
 \mu_0=\gamma \mu   \qquad ;
\end{equation}
both these effects have been pointed out by Israel \cite{israel}.
Finally, the phase-space distribution can be written in the compact form:
\begin{equation}\label{phsp3}
  f({\bf x},{\bf p})_{\tau\sigma} = \lambda \, \e^{-\beta \cdot p} 
  \, \frac{1}{2} \left( D^S([p]^{-1} {\sf R}_{\hat \omegav}(i \omega/T)[p])+
  D^S([p]^{\dagger} {\sf R}_{\hat \omegav}(i \omega/T) [p]^{\dagger-1}) 
  \right)_{\tau \sigma}
\end{equation}
which definitely shows that the ideal rotating gas is made of cells moving at
the four-velocity $u(x)$ pertaining to a rigid motion.

The rotational grand-canonical partition function of the ideal Boltzmann gas can 
be readily calculated by using the single-particle matrix element of 
eq.~(\ref{matel4}) for $\sigma=\tau$, integrating over momenta, summing over 
polarization states and finally summing over all possible particle multiplicities
$N$, dividing by $N!$, as usual. This eventually yields the simple result, in
the approximation (\ref{approx}):
\begin{equation}\label{zomega2}
  Z_\omega = \exp \left[ \int_V \d^3 \x \int \d^3 \p \; f({\bf x},{\bf p}) \right]
  =  \exp \left[ \lambda \, \chiomt
  \int_V \d^3 \x \int \d^3 \p \; \e^{-\beta(x) \cdot p} \right]
\end{equation}
%
 
\section{The ideal rotating gas as a perfect fluid}
\label{fluid}

We have shown in the previous section that the globally equilibrated rotating gas 
can be seen as a set of locally equilibrated cells moving with a rigid velocity field
(\ref{rigid}), at least in the limit $\omega/T \ll 1$. In this section, we will
reinforce this interpretation providing the relevant stress-energy tensor and the
local thermodynamic relation.

The stress-energy tensor $T^{\mu \nu}$ can be calculated directly from particle 
phase space density $f({\bf x},{\bf p})$:
\begin{equation}
 T^{\mu \nu} = \int \frac{\d^3 \p}{\varepsilon} p^{\mu} p^\nu
 f({\bf x},{\bf p})
\end{equation}
thus, for the ideal relativistic Boltzmann gas, according to (\ref{phsp1}), and
using the four-temperature vector $\beta$:
\begin{equation}\label{stress}
  T^{\mu \nu} = \lambda \chiomt \frac{\partial}{\partial \beta^\mu}
  \frac{\partial}{\partial \beta^\nu} \int \frac{\d^3 \p}{\varepsilon} 
  \e^{-\beta \cdot p}
 \end{equation}
The integral on the right-hand side of (\ref{stress}) is apparently Lorentz-invariant, 
therefore it can only depends on $\sqrt{\beta^2}=1/T_0$; if we define it as 
$F(\sqrt{\beta^2})$ we get, after some algebra:
\begin{eqnarray}
 T^{\mu \nu} &=& \lambda \, \chiomt \,
 F''(\sqrt{\beta^2}) \frac{\beta^\mu \beta^\nu}{\beta^2} +\lambda \, \chiomt \,
 F'(\sqrt{\beta^2}) \frac{g^{\mu \nu} \beta^2 - \beta^\mu \beta^\nu}
 {\sqrt{\beta^2}\beta^2} \nonumber \\
 &=& \lambda \, \chiomt \, F''(\sqrt{\beta^2}) u^\mu u^\nu + \lambda \, \chiomt \,
 \frac{F'(\sqrt{\beta^2})}{\sqrt{\beta^2}} (g^{\mu \nu} - u^\mu u^\nu)
\end{eqnarray}
We are therefore led to identify the pressure and the proper energy density
with \footnote{ The first equality in (\ref{prho}) can be easily proved by taking 
the derivative of $F(\sqrt{\beta^2})$ with respect to $\beta^0$, while the second 
is most easily proved by taking into account that
$$
\frac{\beta^\mu}{\sqrt{\beta^2}} \frac{\partial}{\partial \beta^\mu}=
\frac{\d}{\d \sqrt{\beta^2}}
$$}
\begin{eqnarray}\label{prho}
 p &=& -\lambda \, \chiomt \, \frac{F'(\sqrt{\beta^2})}{\sqrt{\beta^2}} =  
 T \int \d^3 \p \; f({\bf x},{\bf p}) \nonumber \\
 \rho &=& \lambda \chiomt F''(\sqrt{\beta^2}) = \int \d^3 \p \; 
 (\varepsilon - {\bf v}\cdot {\bf p}) f({\bf x},{\bf p}) 
\end{eqnarray}
and the stress-energy tensor thereby becomes:
\begin{equation}\label{ideal}
 T^{\mu \nu} = (\rho + p) u^\mu u^\nu - p g^{\mu \nu}
\end{equation}
which is the required form for a fluid at full thermodynamical equilibrium \cite{israel}. 
The identification of pressure with the expression in eq.~(\ref{prho}) dictates that 
$\omega/T = \gamma \omega/T_0$ is a Lorentz invariant quantity; this will be further 
discussed in Sect.~\ref{spin}.

The correctness of the identification of pressure and proper energy density is 
confirmed by the analysis of the basic thermodynamical equation. For a general 
system in the rotational grand-canonical ensemble, entropy can be obtained from 
its definition:
$$
    S = -\tr[ \widehat\rho_\omega \log \widehat\rho_\omega]
$$    
By using the expression of $\widehat\rho_\omega$ in (\ref{denmat2}), we readily
arrive at the basic thermodynamical relation for this ensemble involving the 
expectation values of global quantities:
\begin{equation}\label{global}
  T S = \langle E \rangle - \mu \langle Q \rangle + T \log Z_\omega -
   \omegav \cdot \langle {\bf J} \rangle
\end{equation}
where $T$ is the global temperature. The equation (\ref{global}) can be transformed
into a local relation by taking into account that $\log Z_\omega$ is in fact a 
spacial integral over the system's domain $V$. Comparing (\ref{zomega2}) with
the first of (\ref{prho}) we can write:
\begin{equation}\label{zomega3}
 \log Z_\omega = \frac{1}{T} \int \d^3 \x \; p({\bf x}) 
\end{equation}
This equation has been derived here for the ideal Boltzmann gas, but it holds 
for a general rotating system at thermodynamical equilibrium, as we will see. 
The mean values of energy, charge and angular momentum can be obtained from the 
partition function (\ref{zomega}) by taking derivatives with respect to temperature,
chemical potential and angular velocity. As the operators $\widehat Q$, 
$\widehat H$ and $\widehat {\bf J}$ commute with each other, this gives:
\begin{eqnarray}\label{means}
  && \frac{\partial \log Z_\omega}{\partial T} = \frac{1}{T^2} 
  \left( \langle \widehat H \rangle
   - \mu \langle \widehat Q \rangle + \omegav \cdot \langle \widehat {\bf J} 
    \rangle \right) \nonumber \\
  && \frac{\partial \log Z_\omega}{\partial \mu} = \frac{1}{T} \langle \widehat Q \rangle 
   \nonumber \\
  && \frac{\partial \log Z_\omega}{\partial \omega} = \frac{1}{T} 
   \langle \widehat {\bf J} \cdot \hat\omegav \rangle
\end{eqnarray}
Thus, if $\log Z_\omega$ is an integral over the system's domain of some function,
the same happens for its derivatives and one can then re-write the entropy as
an integral:
\begin{equation}    
 T S = T \int_V \d^3 \x \; \frac{\d S}{\d^3 \x} = 
 \int_V \d^3 {\bf x} \; \frac{\d \langle E \rangle}{\d^3 \x} -
 \mu \frac{\d \langle Q \rangle}{\d^3 \x} + \frac{\d \log Z_\omega}{\d^3 \x}
 - \omegav \cdot \frac{\d \langle {\bf J} \rangle}{\d^3 \x}
\end{equation}
The global temperature $T$ is independent of ${\bf x}$ and one can then assume
that the above relation holds locally:
\begin{equation}\label{local}    
 T \frac{\d S}{\d^3 \x} = 
 \frac{\d \langle E \rangle}{\d^3 \x} -
 \mu \frac{\d \langle Q \rangle}{\d^3 \x} + T \frac{\d \log Z_\omega}{\d^3 \x}
 - \omegav \cdot \frac{\d \langle {\bf J} \rangle}{\d^3 \x}
\end{equation}
In general, the angular momentum density can be written as the sum of an orbital
and a spin part:
\begin{equation}\label{decomp}
  \frac{\d \langle {\bf J} \rangle}{\d^3 \x} = 
  {\bf x} \times \frac{\d \langle {\bf P} \rangle}{\d^3 \x} + 
  \frac{\d \langle {\bf S} \rangle}{\d^3 \x} 
\end{equation}
where $\d \langle {\bf P} \rangle / \d^3 \x$ is the linear momentum density.  
For the present, such decomposition can be explicitely demonstrated, for the ideal
relativistic Boltzmann gas, only for the component of the angular momentum along 
$\omegav$. Indeed, plugging in the eq.~(\ref{means}) the partition function 
(\ref{zomega2}) and using the (\ref{temp}) and (\ref{rigid}):
\begin{equation}\label{jom}
 \langle \widehat {\bf J} \cdot \hat\omegav \rangle = T 
  \frac{\partial \log Z_\omega}{\partial \omega} = 
  \int \d^3 \x \; \hat\omegav \cdot \int \d^3 \p \; \left( {\bf x} \times {\bf p} 
  \right) f({\bf x},{\bf p}) + \frac{\chi'\! \left( \omt \right)}{\chiomt} 
   f({\bf x},{\bf p})
\end{equation}
which allows us to clearly identify the orbital and spin part of the angular momentum
density. The complete calculation of all the components will be presented in 
Sect.~\ref{spin}.

By using the eq.~(\ref{decomp}) and recalling the relations $T=T_0/\gamma$, 
$\mu=\mu_0/\gamma$, the (\ref{local}) can be rewritten as:
\begin{equation}\label{local2}    
 T_0 \frac{1}{\gamma}\frac{\d S}{\d^3 \x} = 
 \gamma \frac{1}{\gamma}\frac{\d \langle E \rangle}{\d^3 \x} -
 \gamma(\omegav \times {\bf x}) \frac{1}{\gamma}\cdot \frac{\d \langle {\bf P} 
 \rangle}{\d^3 \x} - \mu_0 \frac{1}{\gamma} \frac{\d \langle Q \rangle}{\d^3 \x} + 
 T_0 \frac{1}{\gamma}\frac{\d \log Z_\omega}{\d^3 \x}
 - \gamma \omegav \cdot \frac{1}{\gamma}\frac{\d \langle {\bf S} \rangle}{\d^3 \x}
\end{equation}
where we have multiplied and divided by $\gamma=(1-v^2)^{-1}$ where needed.
Now:
$$
  \gamma \frac{\d \langle E \rangle}{\d^3 \x} -
  \gamma(\omegav \times {\bf x}) \cdot \frac{\d \langle {\bf P} 
  \rangle}{\d^3 \x}
$$
is manifestly the energy of the local cell with volume $\d^3 \x$ seen in the cell's
comoving frame, i.e. its {\em mass density}, so that we can also write the (\ref{local2})
as:
\begin{equation}\label{local3}    
 T_0 \frac{1}{\gamma}\frac{\d S}{\d^3 \x} = 
 \frac{1}{\gamma}\frac{\d \langle M \rangle}{\d^3 \x} -
 - \mu_0 \frac{1}{\gamma} \frac{\d \langle Q \rangle}{\d^3 \x} + 
 T_0 \frac{1}{\gamma}\frac{\d \log Z_\omega}{\d^3 \x}
 - \gamma \omegav \cdot \frac{1}{\gamma}\frac{\d \langle {\bf S} \rangle}{\d^3 \x}
\end{equation}
In the eq.~(\ref{local3}) now appear densities rescaled by the Lorentz factor
$\gamma$, i.e. densities measured in the comoving frame or proper densities. We 
can therefore write the above equality in the more familiar form:
\begin{equation}\label{local4}
  T_0 s = \rho - \mu_0 q + p - \gamma \omegav \cdot \sigmav
\end{equation}
where $s$ is the proper entropy density, $\rho$ the proper energy density, $q$ the
proper charge density and $\sigmav$ the {\em proper spin density}. The term 
$T \d \log Z_\omega/\d^3 \x$ has been identified with the pressure on the basis of
(\ref{zomega3}), which has been proved for the ideal Boltzmann gas. Nevertheless, 
since only through this identification the eq.~(\ref{local4}) gets the form of the 
basic local thermodynamical relation, the (\ref{zomega3}) must have a general validity.

With respect to the usually known form, the (\ref{local4}) contains an additional term
owing to the non-vanishing spin density. In the non-relativistic limit this term
becomes $\omegav \cdot \sigmav$ and is in fact responsible for the Barnett effect
\cite{barnett}, i.e. the spontaneous magnetization of a body spun around its axis
when reaching thermodynamical equilibrium \cite{barnett2}. Unlike all other terms, 
the spin-rotation term has a non-covariant form. Expressing it, as necessary, in a 
fully covariant form, requires the introduction of two tensors, the acceleration 
tensor $\Omega$ and the spin density tensor $\sigma$. It will then become clear that 
the general covariant form of (\ref{local4}) is:
\begin{equation}\label{localfin}
  T_0 s = \rho - \mu_0 q + p - \frac{1}{2} \Omega_{\mu \nu} \sigma^{\mu \nu}
\end{equation}
Deriving the tensors $\Omega$ and $\sigma$ will be the subjects of the two forthcoming 
sections.
 
\section{The acceleration tensor}
\label{angvel}

What is the right relativistic generalization of the $\omegav$ vector in a rigid 
velocity field (\ref{rigid})? This is the {\em acceleration tensor} $\Omega$, which 
is discussed for instance in ref.~\cite{misner}, as well as in several papers quoted 
in the bibliography, e.g.~\cite{halb}. For the ease of reading, we spend this 
Section to introduce and describe the main features of this tensor.

The starting point is the definition of an observer and a moving frame in the (Minkowski) 
spacetime, which is meant to be a generalization of an accelerated frame in 
non-relativistic physics. An observer is simply a world-line in spacetime. To this 
world-line one can attach a set of four unit vectors, called a {\em tetrad}, which are a 
basis of the tangent space in each point of the world-line are in fact a reference
frame for the observer. The four tetrad vectors $e_i$ are labelled by indices 
with the convention that the time-like vector $e_0$ is the four-velocity $u$, i.e.
the vector tangent to the world-line. The other three space-like unit vectors can
be chosen at will. The tetrad is orthonormal in the tangent space, thus:
$$
  e_i \cdot e_j = e^\mu_i e_{\mu j} = \eta_{ij} 
$$
being $\eta= {\rm diag} (1,-1,-1,-1)$. Furthermore, any vector of the tangent space
can be decomposed along the tetrad, hence the tetrad is complete:
\begin{equation}\label{complete}
 {\sf I} = \sum_i e_i \otimes e^i \Longrightarrow 
 \sum_i e_{i \mu} \, e^i_\nu = g_{\mu \nu}  
\end{equation}
Since the tetrads at different space-time points of the world-line are orthonormal,
the evolution of a tetrad along the observer world-line can be described by means 
of the Lorentz transformation $\sf \Lambda$ connecting the initial tetrad at the proper 
time $\tau_0$ with that at the proper time $\tau$. Taking advantage of the group
structure of Lorentz transformations, it is also possible to write the tetrad at 
the time $\tau + \d \tau$ as:
$$
  e_i (\tau + \d \tau) = {\sf \Lambda}(\tau + \d \tau)(e_i(0)) = {\sf \Lambda}'
 (\d \tau) {\sf \Lambda}(\tau) ((e_i(0)) = {\sf \Lambda}'(\d \tau) ((e_i(\tau))
$$
where ${\sf \Lambda}'$ is the infinitesimal Lorentz transformation connecting $e_i
(\tau)$ to $e_i(\tau + \d \tau)$. Using the generators of Lorentz group 
${\sf J}_{\mu \nu}$:
\begin{equation}\label{tetrad3}
 e_i (\tau + \d \tau) = \left[{\sf I} - \frac{i}{2} \d\phi^{\mu \nu} 
 {\sf J}_{\mu \nu}  \right] ((e_i(\tau)) = e_i(\tau) - \frac{i}{2} 
 \Omega^{\mu \nu} \d \tau {\sf J}_{\mu \nu}((e_i(\tau))
\end{equation}
where, by definition $\Omega^{\mu \nu} \d \tau$ are the parameters of that 
infinitesimal Lorentz transformation. Reminding that: 
$$
   ({\sf J}_{\mu \nu})^{\alpha}_{\beta} = 
   i \left( \delta^{\alpha}_{\mu} g_{\nu \beta}
   - \delta^{\alpha}_{\nu} g_{\mu \beta} \right)
$$   
and that $\Omega^{\mu \nu}$ is an antisymmetric tensor, we can then rewrite 
(\ref{tetrad3}) as:
\begin{equation}\label{tetrad4}
 \frac{D e_i}{\d \tau}^\alpha \equiv \dot e_i^\alpha = - \frac{i}{2} \Omega^{\mu \nu} 
 ({\sf J}_{\mu \nu})^\alpha_\beta e_i^\beta = \Omega^{\alpha \beta} e_{i \beta}
\end{equation}
or, in a more compact way:
\begin{equation}\label{tetrad5}
  \dot e_i = \Omega \cdot e_i
\end{equation}
Although we are working in a flat spacetime, we emphasized with symbols needed 
for general curved spacetimes, that the derivative in (\ref{tetrad4}) is indeed 
the covariant derivative of the tetrad vector along the world-line, what is also 
denoted with an upper dot. If we are dealing with a fluid, then we will have to 
consider a tetrad field $e_i(x)$, because to each hydrodynamical cell will be 
associated a world-line. In hydrodynamics, the derivative along the cell trajectory 
is called the convective derivative and can be expressed as $u \cdot \partial$.

The tensor $\Omega$ in eq.~(\ref{tetrad4}) is the correct generalization of the 
angular velocity and acceleration and will be henceforth referred to as the 
acceleration tensor, following ref.~\cite{mashhoon}, as has been mentioned. 
The eq.~(\ref{tetrad4}) generalizes the classical equations of motion of the axes 
of a rigid frame in space, i.e.:
\begin{equation}\label{classic}
   \frac{\d \hat {\bf e}}{\d t} = \omegav \times \hat{\bf e}
\end{equation}
We will make the relation between $\Omega^{\mu \nu}$ and $\omegav$ explicit for
the uniform rotating fluid shortly; for the general relation, see Appendix C. 
The equation (\ref{tetrad4}) can be easily inverted to obtain an explicit 
expression of $\Omega^{\mu \nu}$. For this purpose, one has to multiply both sides
of (\ref{tetrad4}) by $e^{i \beta}$, sum over the index $i$ and take advantage of 
the completeness relation (\ref{complete}). The result is:
\begin{equation}\label{omega1}
  \Omega^{\mu \nu} = \sum_i \dot e^{i \mu} e^\nu_i
\end{equation}
In this form, the tensor $\Omega$ is not manifestly antisymmetric, but this can
be readily shown by using again (\ref{complete}):
$$
 \sum_i \dot e^{i \mu} e^\nu_i = \sum_i \dot{\left( e^{i \mu} e^\nu_i \right)}
 - \sum_i e^{i \mu} \dot e^\nu_i = \dot g_{\mu \nu} - 
  \sum_i e^{i \mu} \dot e^\nu_i  = - \sum_i e^{i \mu} \dot e^\nu_i
$$ 

We can now move on to determine the acceleration tensor for the rigidly rotating 
fluid. Yet, before doing this, we face another fundamental question, that is how 
to attach a tetrad to a hydrodynamical cell. When introducing the tetrad, we 
have invoked the world-line of an ``observer", which is always tacitly associated 
with an inner reference frame made of three orthogonal unit vectors, whereas for 
a material fluid pointlike cell, such association is not so straightforward and could 
be done in several manners. For instance, it can be done by using purely geometrical
considerations, like e.g. Fermi-Walker transport or considering the instantaneous 
Lorentz boost without rotation connecting the tetrad to the inertial frame axes; 
or, rather, it can be specified by directions relevant to actual physical quantities 
like e.g. the spin density tensor discussed in the next section. In general, we 
would like to have the most natural definition of the tetrad, possibly determined only 
by the four-velocity $u(x)$ without invoking additional physical fields. Such a 
natural tetrad exists and it is provided by the Frenet-Serret unit vectors 
associated to the world-lines of hydrodynamic cells \cite{freser}, 
i.e. the four-velocity field tangent lines. Even if this definition depends solely 
on the field $u(x)$, it is not unique because, at it is well-known, the four-velocity 
can be associated with the motion of charge (Eckart frame) or energy (Landau frame). 
Notwithstanding, in the present case of full thermodynamical equilibrium, all 
velocities must be the same \cite{israel} and for a macroscopic fluid with finite 
angular momentum are given by (\ref{rigid}). We will later show, in Sect.~\ref{spin},
that this tetrad is precisely the one which makes the spin density tensor, at
least for an ideal Boltzmann gas, proportional to the acceleration tensor. 
Other choices would result in an acceleration tensor not proportional to the
spin density tensor. 
\begin{center}
\begin{figure}[ht]
\epsfxsize=2.5in
\epsffile{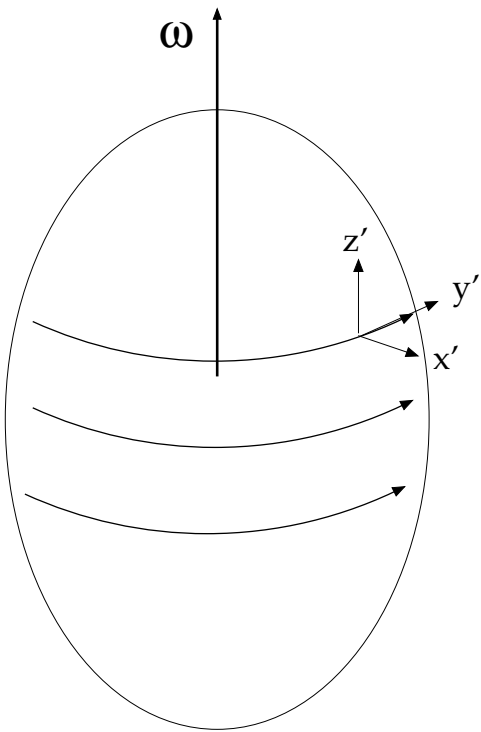} 
\caption{Picture of the rotating gas and the local Frenet-Serret reference frame
or tetrad.
\label{system}}
\end{figure}
\end{center} 

For a uniformly rotating fluid, the Frenet-Serret vectors can be easily determined
and are those corresponding to the axes of a rigidly rotating frame, that is,
in the inertial frame components (see fig.~\ref{system}):
\begin{equation}
 e_0 = u = (\gamma, \gamma {\bf v}) = (\gamma, \gamma v \hat{\bf j}') 
 \qquad e_1 = (0,\hat{\bf i}') \qquad e_2 = (\gamma v, \gamma \hat{\bf j}') \qquad 
 e_3 = (0,\hat{\bf k}')
\end{equation}
whence we can calculate the derivatives along the world-lines. For $i=1,2,3$:
\begin{equation}\label{spacelike}
 \frac{\d e_i}{\d \tau} = \gamma \frac{\d e_i}{\d t} = ( 0, \gamma 
 \omegav \times {\bf e_i})
\end{equation}
because $\gamma$ and $v$ are constant along the field lines. Therefore, the 
Frenet-Serret tetrad is such that the time derivative, in the fixed inertial frame,
of the space vectors yields the simple law (\ref{classic}) involving the assigned
$\omegav$ vector. Taking the derivative of $e_0$, one gets the acceleration 
four-vector $A$:
\begin{equation}\label{timelike}
 \frac{\d e_0}{\d \tau} = \frac{\d u}{\d \tau} \equiv A = 
 \gamma \frac{\d u}{\d t} = (\gamma^4 {\bf v} \cdot {\bf a}, \gamma^4 
 {\bf v} \cdot {\bf a} + \gamma^2 {\bf a}) = (0,\gamma^2{\bf a})
 = (0,\gamma^2 \omegav \times {\bf v})
\end{equation}
where, by definition, ${\bf a} \equiv \d {\bf v}/\d t$ and we have used the explicit
form of the velocity field (\ref{rigid}) implying ${\bf a}= \omegav 
\times {\bf v}$. We are now in a position to calculate the acceleration tensor from 
the formula (\ref{omega1}). The easiest method is to multiply $\Omega$ scalarly on 
both sides by the basis unit vectors of the inertial frame, that is:
\begin{equation}
   \Omega_{\mu \nu} = E_\mu \cdot \sum_i \dot e_i \otimes e^i \cdot E_\nu
\end{equation}
which leads us to, being $\omegav = \omega \hat {\bf k} = \omega \hat {\bf k}'$,
and using eqs.~(\ref{timelike}) and (\ref{spacelike}):
\begin{equation}\label{omegatens}
 \Omega = 
 \left( \begin{array}{cccc} 
                     0 & 0 & 0 & 0  \\
                     0 & 0 & \gamma \omega & 0 \\
		     0 & -\gamma \omega& 0 & 0 \\
		     0 & 0 & 0 & 0 
  \end{array} \right)
\end{equation}
It is worth pointing out that the double contraction of $\Omega$ with itself 
yields:
\begin{equation}\label{contrac}
 \Omega : \Omega \equiv \Omega_{\mu \nu} \Omega^{\mu \nu} = 
 2 \gamma^2 \omega^2
\end{equation} 
and, therefore, $\gamma \omega$ must be a Lorentz invariant; this is further 
illustrated in Appendix C. If $\gamma \omega$ is a Lorentz invariant, so is the 
ratio $\omega/T$ as:
\begin{equation}\label{omegat}
  \omt = \frac{\omega \gamma}{\gamma T} = \frac{\sqrt{\frac{1}{2}\Omega:\Omega}}{T_0}
   = \sqrt{\frac{1}{2} \, \beta^2 \, \Omega:\Omega}
\end{equation}
As a consequence, pressure in eq.~(\ref{prho}) can be written in a manifestly
covariant form:
\begin{equation}\label{pressure}
 p = \lambda \, \frac{F' (\sqrt{\beta^2})}{\sqrt{\beta^2}} 
 \chi \! \left( \sqrt {\frac{1}{2} \, \beta^2 \, \Omega : \Omega} \right) 
\end{equation}
as well as all other local thermodynamical quantities.

\section{The spin tensor}
\label{spin}

We now come to the central subject of this work, i.e. the determination of the spin
density tensor for the ideal rotating gas and its relation with the acceleration
tensor. Some preliminary definitions are necessary. For a fluid, angular momentum 
conservation is expressed by an equation like:
\begin{equation}
 \partial_\lambda {\cal J}^{\lambda, \mu \nu} = \partial_\lambda 
 ({\cal S}^{\lambda, \mu \nu} + x^\mu T^{\lambda \nu} - x^\mu T^{\lambda \mu})
 = 0 
\end{equation} 
where ${\cal S}^{\lambda, \mu \nu}$ is a rank 3 tensor antisymmetric in the $\mu \nu$
indices commonly known as {\em spin tensor}. In classical field theories, the spin
tensor can be made vanishing by a suitable redefinition of the stress-energy tensor,
well known as Belinfante symmetrization procedure. However, such procedure does not leave the 
thermodynamic relation (\ref{localfin}) invariant even when $\Omega=0$, because both 
energy density and pressure are affected by the Belinfante transformation (see Appendix 
D for a detailed calculation) and the transformed entropy would have a different local 
and global value, which is unacceptable. The profound reason of the lack of the freedom
of changing spin tensor and stress-energy tensor in presence of thermodynamics, is the 
fact that entropy is an effective counting of degrees of freedom, a concept tightly related 
to the discrete and quantum nature of matter and definitely outside a purely classical 
field mechanical framework.

At full thermodynamical equilibrium we must have a symmetric stress-energy tensor in 
the form (\ref{ideal}) and therefore the spin tensor must be conserved as well:
$$
 \partial_\lambda {\cal S}^{\lambda, \mu \nu} = 0
$$
Furthermore, to meet the requirement that all currents must be parallel at full
thermodynamical equilibrium, the spin tensor must be factorizable into an antisymmetric 
{\em spin density tensor} $\sigma$ and the four-velocity:
\begin{equation}
 {\cal S}^{\lambda, \mu \nu} = \sigma^{\mu \nu} u^\lambda
\end{equation}

To determine the spin density tensor for a rotating ideal Boltzmann gas, it is 
necessary to calculate total angular momentum density and subtract the orbital part. 
A separation between angular and spin part has already been encountered in 
eq.~(\ref{jom}) for the component of the total angular momentum along $\omegav$. 
The orbital angular momentum density {\em must} look like:
$$
  \frac{\d {\bf L}}{\d ^3 \x} = 
  \int \d^3 \p \; ({\bf x} \times {\bf p}) f({\bf x},{\bf p})
$$
hence, if such a form is retrieved in the calculation of the total angular momentum,
the rest can be unambiguously identified as the spin density. In the particular case 
of the Boltzmann ideal gas, particles can be handled independently of each other. This 
makes it easy to calculate the spin density tensor, what can be done by simply 
summing up single particle contributions. We then need to calculate the expectation 
value of the angular momentum tensor (i.e. the Lorentz group generators) over single 
particle states, that is:
\begin{equation}\label{jtens1}
 \langle\langle \widehat j^{\mu \nu} \rangle\rangle \equiv 
 \frac{1}{\langle N \rangle} \sum_{\sigma} \int \d^3 \p \; \bra{p,\sigma} 
 \widehat j^{\mu \nu} \exp[(-\widehat h + \mu \widehat q + \omegav \cdot \widehat 
 {\bf j})/T] \Pro_V \ket{p, \sigma} 
\end{equation}
where the normalizing factor
$$
 \langle N \rangle = \int_V \d^3 \x \int \d^3 \p \; f({\bf x},{\bf p})
$$
concides with the mean number of particles (see Sect.~\ref{eqgas}).
We will proceed separately for the space components of the angular momentum tensor, 
i.e. the properly called angular momentum ${\bf J}$, and the time components, i.e. 
the boost generators ${\sf K}^i={\sf J}^{0i}$. Taking the $z$ axis as the $\hat\omegav$ 
direction, we can write the mean values in (\ref{jtens1}) as derivatives:
\begin{eqnarray}\label{jtens2}
 \langle\langle \widehat j_l \rangle\rangle &=& 
 \frac{1}{\langle N \rangle}
 \sum_\sigma \int \d^3 \p \; i \frac{\partial}{\partial \phi_l} \bra{p,\sigma} 
 \widehat{\sf R}_{\hat{\bf e}_l}(\phi_l) \exp[(-\widehat h + \mu \widehat q + \omega 
 \widehat j_z)/T] \Pro_V \ket{p, \sigma} \Big|_{\phi_l=0} \nonumber \\
 \langle\langle \widehat k_l \rangle\rangle &=& 
 \frac{1}{\langle N \rangle}
 \sum_\sigma \int \d^3 \p \; i \frac{\partial}{\partial \xi_l} \bra{p,\sigma} 
 \widehat {\sf L}_{\hat{\bf e}_l} (\xi_l) 
 \exp[(-\widehat h + \mu \widehat q + \omega \widehat j_z)/T] \Pro_V 
 \ket{p, \sigma} \Big|_{\xi_l=0}
\end{eqnarray}
where ${\sf R}_{\hat{\bf e}_l}(\phi_l) = \exp[-i \phi_l {\sf J}_l]$ is a rotation 
of axis $\hat{\bf e}_l$ and angle $\phi_l$ and ${\sf L}_{\hat{\bf e}_l} (\xi_l)=
\exp[-i \xi_l {\sf K}_l]$ is a Lorentz boost of direction $\hat{\bf e}_l$ and 
hyperbolic angle $\xi_l$. The matrix elements in the above equation can be 
calculated in a similar fashion as for that in eq.~(\ref{matel1}). One just needs 
to insert a resolution of the identity made of momentum-spin eigenstates between 
the operators $\widehat{\sf R}_{\hat{\bf e}_l}(\phi_l),\widehat {\sf L}_{\hat{\bf e}_l} 
(\xi_l)$ and the density operator and work out the obtained expression following 
the procedure drawn in Sect.~\ref{eqgas} from eq.~(\ref{matel1}) to 
(\ref{matel4}). This leads to:
\begin{eqnarray}\label{rotaz}
 \langle\langle \widehat j_l \rangle\rangle &=& 
 \frac{1}{\langle N \rangle}
 \sum_\sigma \int \d^3 \p \; \e^{-\varepsilon/T + \mu q/T} 
 i \frac{\partial}{\partial \phi_l} \int_V \d^3 \x \; 
 \e^{i {\bf x} \cdot ({\bf p}-{\sf R}_{\hat \omegav}(i \omega/T)^{-1}
 {\sf R}_{\hat{\bf e}_l}(\phi_l)^{-1}({\bf p}))} \nonumber \\
 &\times& \frac{1}{2} \left( D^S([p]^{-1} 
 {\sf R}_{\hat{\bf e}_l}(\phi_l){\sf R}_{\hat \omegav}(i \omega/T)[p])+D^S([p]^{\dagger}
 {\sf R}_{\hat{\bf e}_l}(\phi_l){\sf R}_{\hat \omegav}(i \omega/T) [p]^{\dagger-1}) 
 \right)_{\sigma \sigma}\Big|_{\phi_l=0} \nonumber \\
 &=& \frac{1}{\langle N \rangle} \int \d^3 \p \; \e^{-\varepsilon/T + \mu q/T} 
 i \frac{\partial}{\partial \phi_l} \int_V \d^3 \x \; 
 \e^{i {\bf x} \cdot ({\bf p}-{\sf R}_{\hat \omegav}(i \omega/T)^{-1}
 {\sf R}_{\hat{\bf e}_l}(\phi_l)^{-1}({\bf p}))} \tr 
  D^S({\sf R}_{\hat{\bf e}_l}(\phi_l){\sf R}_{\hat \omegav}(i \omega/T))
 \Big|_{\phi_l=0}
\end{eqnarray} 
and
\begin{eqnarray}\label{boost}
 \langle\langle \widehat k_l \rangle\rangle &=& 
 i \frac{\partial}{\partial \xi_l} \frac{1}{\langle N \rangle} \sum_\sigma \int \d^3 \p \; 
 \e^{-{\sf L}_{\hat{\bf e}_l}(\xi_l)^{-1} (p)^0/T + \mu q/T} 
 \sqrt{\frac{\varepsilon}{{\sf L}_{\hat{\bf e}_l}(\xi_l)^{-1}(p)^0}} 
 \int_V \d^3 \x \; \e^{i \sum_{r=1}^3 x^r (p-{\sf R}_{\hat \omegav}(i \omega/T)^{-1}
 {\sf L}_{\hat{\bf e}_l}(\xi_l)^{-1} (p))^r} \nonumber \\
 &\times& \frac{1}{2} \left( D^S([p]^{-1} {\sf L}_{\hat{\bf e}_l}(\xi_l)
 {\sf R}_{\hat \omegav}(i \omega/T)[p])+ D^S([p]^{\dagger}
 {\sf L}_{\hat{\bf e}_l}(\xi_l)^{\dagger -1}{\sf R}_{\hat \omegav}(i \omega/T)
 [p]^{\dagger-1}) \right)_{\sigma \sigma}\Big|_{\xi_l=0} \nonumber \\
 &=& \frac{1}{\langle N \rangle} \int \d^3 \p \; 
 i \frac{\partial}{\partial \xi_l} \e^{-{\sf L}_{\hat{\bf e}_l}(\xi_l)^{-1} (p)^0/T + \mu q/T} 
 \sqrt{\frac{\varepsilon}{{\sf L}_{\hat{\bf e}_l}(\xi_l)^{-1}(p)^0}} 
 \int_V \d^3 \x \; \e^{i \sum_{r=1}^3 x^r (p-{\sf R}_{\hat \omegav}(i \omega/T)^{-1}
 {\sf L}_{\hat{\bf e}_l}(\xi_l)^{-1} (p))^r} \nonumber \\
 &\times& \frac{1}{2} \left[ \tr D^S({\sf L}_{\hat{\bf e}_l}(\xi_l)
 {\sf R}_{\hat \omegav}(i \omega/T)) + \tr D^S({\sf L}_{\hat{\bf e}_l}
 (\xi_l)^{\dagger -1}{\sf R}_{\hat \omegav}(i \omega/T)) \right] \Big|_{\xi_l=0}
\end{eqnarray}
Note the difference between the traces of the finite dimensional representation
matrices in (\ref{rotaz}) and (\ref{boost}) which is owing to the non-unitarity
of the SL(2,C) representation of boosts, i.e. $D^S({\sf L}(\xi))^{\dagger -1} 
\ne D^S({\sf L}(\xi))$.
 
Let us first focus on the eq.~(\ref{rotaz}). If we let ${\bf p'} \equiv 
{\sf R}_{\hat{\bf e}_l}(\phi_l)^{-1}({\bf p})$, the energy as well as the momentum 
integral measure does not change and the derivative with respect to $\phi_l$ can 
be worked out:
\begin{eqnarray}\label{rota2}
 \langle\langle \widehat j_l \rangle\rangle &=& 
 \frac{1}{\langle N \rangle} \int \d^3 \p' \; \e^{-\varepsilon'/T + \mu q/T} 
 i \frac{\partial}{\partial \phi_l} \int_V \d^3 \x \; 
 \e^{i {\bf x} \cdot ({\sf R}_{\hat{\bf e}_l}(\phi_l)({\bf p}'-
 {\sf R}_{\hat \omegav}(i \omega/T)^{-1}({\bf p}'))} 
 \tr D^S({\sf R}_{\hat{\bf e}_l}(\phi_l){\sf R}_{\hat \omegav}(i \omega/T))
 \Big|_{\phi_l=0} \nonumber \\
 &=& \frac{1}{\langle N \rangle} \int \d^3 \p' \; \e^{-\varepsilon'/T + \mu q/T} 
 \int_V \d^3 \x \; \e^{i {\bf x} \cdot ({\bf p}')-{\sf R}_{\hat \omegav}(i \omega/T)^{-1}
 ({\bf p}'))} \left[ ({\bf x} \times {\bf p}') \cdot \hat{\bf e}_l \, \chiomt + 
 \tr D^S({\sf J}_l{\sf R}_{\hat \omegav}(i \omega/T)) \right]
\end{eqnarray} 
where we used the (\ref{chiomegat}). The (\ref{rota2}) can be rewritten in a more
compact form recalling the calculations of Sect.~2:  
\begin{equation}\label{rota3}
 \langle\langle \widehat{\bf j} \rangle\rangle = 
 \frac{1}{\langle N \rangle} \int_V \d^3 \x \; \int \d^3 \p \; 
 f({\bf x},{\bf p}) \left[ ({\bf x} \times {\bf p}) 
 +\frac{1}{\chiomt} \tr \left[D^S({\bf J}{\sf R}_{\hat \omegav}(i \omega/T))\right] 
 \right]
\end{equation} 
where we have redefined ${\bf p'}$ as ${\bf p}$. In this form, eq.~(\ref{rota3}) allows
us to pin down the spacial part of the spin density tensor which is the second term in
the integrand of (\ref{rota3}) and more specifically:
\begin{equation}\label{sdens1}
 \frac{\d {\bf S}}{\d^3 \x} = 
 \int \d^3 \p \; f({\bf x},{\bf p}) \frac{1}{\chiomt} 
 \tr D^S({\bf J}{\sf R}_{\hat \omegav}(i \omega/T)) 
\end{equation}
The $z$ component of the trace is readily calculated as:
\begin{equation}\label{zcomp}
 \tr D^S({\sf J}_z{\sf R}_{\hat \omegav}(i \omega/T)) = 
 \frac{\partial}{\partial \omega/T} \tr D^S({\sf R}_{\hat \omegav}(i \omega/T)) = 
 \frac{\partial}{\partial \omega/T} \tr \exp[\omega D^S({\sf J_z})/T] = 
 \chi'\! \left( \omt \right)
\end{equation}
As far as the $x,y$ components are concerned, we first note that for any analytic
function $f({\sf J})$ of the rotation generators, or a representation thereof, and 
for any operator ${\sf O}$ such that $[{\sf O},{\sf J}_z]=0$, we have:
\begin{equation}\label{vanishtrace}
  \tr \left({\sf J}_{x,y} f({\sf J}_z) {\sf O} \right)=0
\end{equation}
To prove it, we just calculate the trace of ${\sf J}_{x,y} {\sf J}_z^{N+1} {\sf O}$:
taking advantage of the ciclicity of trace and the Lie algebra properties:
\begin{eqnarray}\label{ragiona}
 &&\tr \left(  {\sf J}_{x,y} {\sf J}_z^{N+1} {\sf O} \right) = 
 \tr \left( {\sf J}_{x,y} {\sf J}_z^{N} {\sf O} {\sf J}_z \right) = 
 \tr \left( {\sf J}_z  {\sf J}_{x,y}  {\sf J}_z^{N} {\sf O} \right) \nonumber \\
 && = \tr \left( {\sf J}_{x,y} {\sf J}_z  {\sf J}_z^{N} {\sf O} \right) \pm i \tr 
  \left( {\sf J}_{y,x} {\sf J}_z^{N} {\sf O} \right) = \tr \left( 
   {\sf J}_{x,y}  {\sf J}_z^{N+1} {\sf O} \right) \pm i 
  \tr \left( {\sf J}_{y,x} {\sf J}_z^{N} {\sf O} \right) 
\end{eqnarray}
Comparing the first and last side of the above equation, we conclude that $\tr [
{\sf J}_{y,x} {\sf J}_z^{N} {\sf O}] = 0$ for any $N$ and therefore the 
(\ref{vanishtrace}). We now apply the above reasoning to our specific case where
$f({\sf J}_z) = \exp [ \omega {\sf J}_z/T]$ and ${\sf O} = {\sf I}$, which both 
fulfill the aforementioned conditions and thus conclude that:
\begin{equation}\label{vanishjxy}
 \tr D^S({\sf J}_{x,y}{\sf R}_{\hat \omegav}(i \omega/T)) = 0 
\end{equation}
As a consequence, the spin density (\ref{sdens1}) turns out to be, in view of (\ref{zcomp})
and (\ref{vanishjxy}):
\begin{equation}\label{sdens2}
 \frac{\d {\bf S}}{\d^3 \x} = \int \d^3 \p \; f({\bf x},{\bf p}) 
 \frac{\chi'\! \left( \omt \right)}{\chiomt} \hat\omegav = n({\bf x})
 \frac{\chi'\! \left( \omt \right)}{\chiomt} \hat\omegav
\end{equation}
being $n({\bf x}) = \int \d^3 \p \; f({\bf x},{\bf p})$ the particle density. 
Similarly, the proper spin density (see Section~\ref{fluid}) turns out to be:
\begin{equation}\label{sdens3}
 \sigmav = \frac{1}{\gamma} \frac{\d {\bf S}}{\d^3 \x} =
 \frac{1}{\gamma} n({\bf x}) \frac{\chi'\! \left( \omt \right)}{\chiomt} 
 \hat\omegav = \nu \frac{\chi'\! \left( \omt \right)}{\chiomt} 
 \hat\omegav 
\end{equation}
being $\nu$ the proper particle density:
\begin{equation}\label{equstate}
  \nu = \frac{1}{\gamma} n({\bf x}) = \frac{1}{\gamma} 
  \int \d^3 \p \; f({\bf x},{\bf p}) = \frac{T}{T_0} 
  \int \d^3 \p \; f({\bf x},{\bf p}) = \frac{p}{T_0}
\end{equation}
where we have used the equation (\ref{prho}).
We note in passing that the $x,y$ components of the orbital part of the angular 
momentum density vanish as well because the system (and the phase space distribution)
is geometrically invariant by rotation around the $z$ axis.  

We can now turn to the eq.~(\ref{boost}). We have seen, for angular momentum, that 
the derivative with respect to group parameter produces two terms: the first, obtained 
by taking the derivative of the spacial integrand, gives rise to an ``orbital" part, 
while the second term stems from the derivative of the matrices of the 
finite-dimensional SL(2,C) representation and gives rise to the actual ``spin" part. 
Similarly, for boosts, the derivative of SL(2,C) representation matrices yields the 
contribution to the spin tensor density while the derivative of the spacial integral
generates an orbital part:
\begin{eqnarray}\label{orboost1}
 \langle\langle \widehat k_l \rangle\rangle_{\rm orb} &=& 
 \frac{1}{\langle N \rangle} \lambda \chiomt \int \d^3 \p \; 
 i \frac{\partial}{\partial \xi_l} \e^{-{\sf L}_{\hat{\bf e}_l}(\xi_l)^{-1} (p)^0/T} 
 \sqrt{\frac{\varepsilon}{{\sf L}_{\hat{\bf e}_l}(\xi_l)^{-1}(p)^0}} 
 \int_V \d^3 \x \; \e^{i \sum_{r=1}^3 x^r (p-{\sf R}_{\hat \omegav}(i \omega/T)^{-1}
 {\sf L}_{\hat{\bf e}_l}(\xi_l)^{-1} (p))^r} \Big|_{\xi_l=0} \nonumber \\
 &=& \frac{1}{\langle N \rangle} \lambda \chiomt 
 \int \frac{\d^3 \p}{\varepsilon} \; \varepsilon \, i \frac{\partial}{\partial \xi_l}
 \e^{-{\sf L}_{\hat{\bf e}_l}(\xi_l)^{-1}(p)^0/T}  
 \sqrt{\frac{\varepsilon}{{\sf L}_{\hat{\bf e}_l}(\xi_l)^{-1}(p)^0}} 
 \int_V \d^3 \x \; \e^{i \sum_{r=1}^3 x^r (p-{\sf R}_{\hat \omegav}(i \omega/T)^{-1}
 {\sf L}_{\hat{\bf e}_l}(\xi_l)^{-1} (p))^r} \Big|_{\xi_l=0}
\end{eqnarray}
To go on, we need to work out the action of a Lorentz boost of hyperbolic angle $\xi$ 
along the direction $\hat{\bf n}$ on the four-momentum $p$:
\begin{eqnarray}\label{lboost}
 && {\sf L}_{\hat{\bf n}}(\xi)(p)^0 = p^0 \cosh \xi  + {\hat{\bf n}}
  \cdot {\bf p} \sinh \xi  \nonumber \\
 && {\sf L}_{\hat{\bf n}}(\xi)(p)^i = p^i + p^0 \sinh \xi {\hat n}^i +
 (\cosh \xi - 1) {\hat{\bf n}} \cdot {\bf p} \, {\hat n}^i
\end{eqnarray}
Let $p'= {\sf L}_{\hat{\bf e}_l}(\xi_l)^{-1} (p)$ and implement this change of variable
in the integral of eq.~(\ref{orboost1}):
\begin{equation}\label{orboost2}
 \langle\langle \widehat k_l \rangle\rangle_{\rm orb} =
 \frac{1}{\langle N \rangle} \lambda \chiomt i \frac{\partial}{\partial \xi_l}
 \int \frac{\d^3 \p'}{\varepsilon'}\; \e^{-\varepsilon'/T} \,
 [{\sf L}_{\hat{\bf e}_l}(\xi_l)(p')^0]^{3/2}\frac{1}{\sqrt{\varepsilon'}} 
 \int_V \d^3 \x \; \e^{i \sum_{r=1}^3 x^r ({\sf L}_{\hat{\bf e}_l}(\xi_l)(p')
 -{\sf R}_{\hat \omegav}(i \omega/T)^{-1}(p'))^r} \Big|_{\xi_l=0}
\end{equation}
The derivative can be worked out by using (\ref{lboost}):
\begin{eqnarray}\label{orboost3}
 \langle\langle \widehat k_l \rangle\rangle_{\rm orb} &=& 
 \frac{1}{\langle N \rangle} \lambda \chiomt i \frac{\partial}{\partial \xi_l}
 \int \d^3 \p' \; \frac{[{\sf L}_{\hat{\bf e}_l}(\xi_l)(p')^0]^{3/2}}
 {\varepsilon'^{3/2}} \,  \e^{-\varepsilon'/T}  
 \int_V \d^3 \x \; \e^{i \sum_{r=1}^3 x^r ({\sf L}_{\hat{\bf e}_l}(\xi_l)(p')
 -{\sf R}_{\hat \omegav}(i \omega/T)^{-1}(p'))^r} \Big|_{\xi_l=0}
 \nonumber \\
 &=& \frac{1}{\langle N \rangle} \lambda \chiomt \int \d^3 \p' \; \e^{-\varepsilon'/T} 
 \int_V \d^3 \x \; \left[\frac{3i}{2 \varepsilon'} 
 {\bf p}'\cdot{\hat{\bf e}}_l - \varepsilon' {\bf x} \cdot {\hat{\bf e}}_l \right]
 \e^{i {\bf x} \cdot ({\bf p}'-{\sf R}_{\hat \omegav}(i \omega/T)^{-1}({\bf p}'))}
 \nonumber \\ 
 &=& \frac{1}{\langle N \rangle}\int_V \d^3 \x \int \d^3 \p \;
 \left[ \frac{3i}{2 \varepsilon} 
 {\bf p} \cdot{\hat{\bf e}}_l - \varepsilon \, {\bf x}\cdot {\hat{\bf e}}_l \right]  
 f({\bf x},{\bf p})
\end{eqnarray}
where, in the last expression, we have redefined ${\bf p}'$ as ${\bf p}$.
It is now easy to realize that, in a rotating system, the last integral vanishes.
In fact, we have:
\begin{equation}\label{orboost4}
 \frac{\d \langle\langle \widehat {\bf k} \rangle\rangle_{\rm orb}}{\d^3 \p} 
 = \frac{1}{\langle N \rangle} \frac{3i}{2 \varepsilon} {\bf p} \int_V \d^3 \x \; 
 f({\bf x},{\bf p}) - \frac{1}{\langle N \rangle}
 \varepsilon \, \int_V \d^3 \x \; {\bf x} f({\bf x},{\bf p})
\end{equation}
The rightmost integral yields the origin of the inertial frame, which is centered 
on the axis of the rotating system according to the (\ref{rigid}). The leftmost 
integral, on the other hand, is imaginary and the integrand is proportional 
to particle velocity; once integrated over momentum, this term yields the mean 
particle velocity which vanishes. Thus, for any constant $C$:
$$
  \langle\langle \widehat{\bf k} \rangle\rangle_{\rm{orb}} = \frac{1}{\langle N \rangle}
  \int_V \d^3 \x \; \int \d^3 \p \; (C \frac{{\bf p}}{\varepsilon} - 
  \varepsilon {\bf x}) f({\bf x},{\bf p}) = 0
$$. 

On the other hand, the ``spin" part of the eq.~(\ref{boost}) can be written as:
\begin{eqnarray}\label{kboost}
 \langle\langle \widehat k_l \rangle\rangle_{\rm{spin}} &=& 
 \frac{1}{\langle N \rangle} \int \d^3 \p \; 
 \e^{-\varepsilon/T + \mu q/T} \int_V \d^3 \x \; \e^{i  {\bf x} \cdot 
 ({\bf p} -{\sf R}_{\hat \omegav} (i \omega/T)^{-1}({\bf p}))} \nonumber \\
 &\times& i \frac{\partial}{\partial \xi}
 \frac{1}{2} \left( \tr \left[ D^S({\sf L}_{\hat{\bf e}_l}(\xi_l)
 {\sf R}_{\hat \omegav}(i \omega/T))\right] + \tr \left[ D^S(
 {\sf L}_{\hat{\bf e}_l}(\xi_l)^{\dagger -1}{\sf R}_{\hat \omegav}(i \omega/T)) 
 \right] \right)\Big|_{\xi_l=0} \nonumber \\
 &=& \frac{1}{\langle N \rangle} \int_V \d^3 \x \int \d^3 \p \; 
  f({\bf x},{\bf p}) \frac{1}{2} \tr \left[ D^S({\sf K}_l {\sf R}_{\hat \omegav}
  (i \omega/T))+ D^S({\sf K}^\dagger_l {\sf R}_{\hat \omegav}(i \omega/T)) 
  \right]
\end{eqnarray}
By using the known properties of representations, the last matrix can be written 
as: 
$$ 
D^S({\sf K}_l){\sf R}_{\hat \omegav}(i \omega/T))+ 
D^S({\sf K}^\dagger_l {\sf R}_{\hat \omegav}(i \omega/T)) = 
D^S({\sf K}_l)D^S({\sf R}_{\hat \omegav}(i \omega/T))+ 
D^S({\sf K}^\dagger_l) D^S({\sf R}_{\hat \omegav}(i \omega/T)) 
$$
but in fact, for finite-dimensional SL(2,C) representations $D^S({\sf K})$ is 
anti-hermitian, hence $D^S({\sf K}^\dagger)= - D^S({\sf K})$ and the above expression 
vanishes. Consequently, the contribution of boosts to spin density in a rotating
ideal Boltzmann gas is zero, i.e.: 
\begin{equation}\label{kspin}
 \frac{\d S^{0i}}{\d^3 \x} = 0 \qquad {\rm or} 
 \qquad \frac{\d {\bf K}_{\rm spin}}{\d^3 \x} = 0
\end{equation}
Collecting the eqs.~(\ref{sdens3}) and (\ref{kspin}) we can finally write the spin
density tensor $\sigma$ as:
\begin{equation}\label{sigmatens}
 \sigma_{\mu \nu} = 
 \left( \begin{array}{cccc} 
    0 & 0 & 0 & 0  \\
    0 & 0 & \nu \; \frac{\chi'\! \left( \omt \right)}{\chiomt} & 0 \\
    0 & -\nu \; \frac{\chi'\! \left( \omt \right)}{\chiomt} & 0 & 0 \\
    0 & 0 & 0 & 0 
  \end{array} \right)
\end{equation}

The result (\ref{sigmatens}) has three major implications. Firstly, for an ideal 
relativistic Boltzmann gas at thermodynamical equilibrium, the spin density tensor 
(\ref{sigmatens}) is proportional to the acceleration tensor (\ref{omegatens}) 
defined with the Frenet-Serret tetrad, through a Lorentz scalar $\iota$ depending 
on the intensive invariant parameters $T_0$, $\mu_0$ and $\Omega:\Omega$:
\begin{equation}\label{sigmarel}
 \sigma_{\mu \nu} = \iota \Omega_{\mu \nu} 
\end{equation}
with:
\begin{equation}\label{iotadef} 
  \iota = \nu \frac{1}{\gamma \omega} \; \frac{\chi'\! \left( \omt \right)}{\chiomt} 
 = \nu \frac{1}{\sqrt {\frac{1}{2} \Omega : \Omega}} = \frac{p}{T_0} 
 \frac{1}{\sqrt {\frac{1}{2} \Omega : \Omega}} =
 \lambda \, \chi'\! \left( \sqrt {\frac{1}{2} \, \beta^2 \, \Omega : \Omega} \right) 
 F'(\sqrt{\beta^2})
\end{equation}
where, in (\ref{iotadef}), we have used the eqs.~(\ref{contrac}), (\ref{omegat}),
(\ref{equstate}) and (\ref{pressure}). The last form of $\iota$ emphasizes its 
nature of Lorentz scalar depending on the Lorentz scalars $\mu_0$, $\beta^2$ and 
$\Omega:\Omega$. Also, since $\omega \ll T$ and using the (\ref{chiomegat}):
\begin{equation}
 \iota \simeq \frac{\nu}{\gamma \omega} \frac{S(S+1)}{3} \omt = 
 \frac{\nu}{\gamma T} \frac{S(S+1)}{3} = \frac{\nu}{T_0} \frac{S(S+1)}{3}
 \qquad \Longrightarrow \iota \simeq \hbar^2 \frac{\nu}{K T_0} \frac{S(S+1)}{3}
\end{equation} 
where in the last expression we have restored the natural constants.

Secondly, the basic thermodynamical relation (\ref{local4}) can be written as:
\begin{equation}\label{thermo}
 T_0 s = \rho - \mu_0 q + p - \frac{1}{2} \, \Omega : \sigma =
 \rho - \mu_0 q + p - \frac{1}{2} \, \Omega_{\mu \nu} \sigma^{\mu \nu}
\end{equation}  
as anticipated in eq.~(\ref{localfin}). For an ideal Boltzmann gas the additional
term is:
\begin{equation}\label{term}
 \frac{1}{2} \, \Omega : \sigma = \frac{1}{2} \, \iota \Omega : \Omega =
 \nu \sqrt {\frac{1}{2} \Omega : \Omega} \; 
 \frac{\chi'\! \left( \omt \right)}{\chiomt}  \simeq \nu \gamma 
 \frac{\omega^2}{T} \frac{S(S+1)}{3} \qquad \Longrightarrow 
 \nu \gamma \frac{\hbar^2 \omega^2}{KT} \frac{S(S+1)}{3} 
\end{equation}
using (\ref{iotadef}). Once again, in the eq.~(\ref{term}), we have restored the
natural units, which allows us to conclude that for ordinary macroscopic gases
the correction to thermodynamic relation owing to spin is extremely small, being
$\hbar \omega/KT \ll 1$ under all circumstances. For instance, for a mole of a 
monoatomic gas with $S=1$ at 273 $^{\rm o}$K, $p = 10^5$ Pa and $\omega = 100$ Hz, 
$\hbar \omega/KT = {\cal O}(10^{-12})$ and the energy density associated to 
spin-angular velocity term is of the order of $10^{-19}$J/m$^3$. However, in 
extreme situation, like e.g. in astrophysics or relativistic heavy ion collisions, 
this term could be much more important. Also, as has been mentioned at the end of 
Sect.~\ref{fluid}, the term coupling spin and angular velocity-acceleration is 
responsible for the magnetization observed in rotating bodies, the Barnett effect 
\cite{barnett,barnett2}. 

Finally, the projection of the spin density tensor over the four-velocity field,
is non-vanishing and proportional to the four-acceleration $A$ through the factor
$\iota$:
\begin{equation}\label{tdef}
  t^{\mu} \equiv \sigma^{\mu \nu} u_\nu = \iota \, \Omega^{\mu \nu} u_\nu
  = \iota \, A^\mu
\end{equation}    
Particularly, in the inertial frame, this vector has time and space components:
\begin{equation}\label{tvect}
   t = (0, \iota \gamma^2 \omegav \times {\bf v})
\end{equation}
(see eq.~(\ref{timelike}) and it is thus orthogonal to both $\omegav$ and the 
velocity. This result is in contrast with common assumptions in the hydrodynamics
of fluids with spin, as has been mentioned in the Introduction, but it is not 
surprising as there is no principle reason for it to vanish in a system of many
particles \cite{hagedorn}. Indeed, while the vector $t$ for a quantum relativistic
particle in a pure state vanishes, this is generally neither the case for a 
many-particle states nor for a single particle in a mixture of states (see 
Appendix E). 

We end this Section by noting that the calculations herein performed provide an 
alternative way of determining the polarization four-vector of particles, which was
computed explicitely for different spins in ref.~\cite{becapicc}. The relevant
proof is shown in Appendix F.

\section{Viewpoint of the accelerated observer}
\label{accel}

Since the ideal gas we are dealing with is a macroscopic system, we can ideally 
attach to each point a physical observer (i.e. a tetrad) moving along the velocity 
flow lines. For him to observe local thermodynamical equilibrium, it is necessary
that all (local) physical quantities are constant in his frame. If this is the
case, and assuming the locality hypothesis \cite{longhi,mashhoon}, stating that 
the accelerated observer measures the same quantities as the comoving observer 
(i.e. the inertial observer instantaneously coinciding with the accelerated one),
the relation he finds among the local thermodynamical quantities are actual 
equations of state. Particularly, the relation (\ref{sigmarel}) linking a
kinematical quantity like the acceleration tensor and the spin density tensor
is to be promoted to the rank of equation of state for a system in local thermodynamical 
equilibrium. 

The condition of steadiness for the non-inertial observer means, for a scalar:
$$
  \frac{\d X}{\d \tau} = 0
$$
where $\tau$ is the proper time. For vector quantities, this turns into a 
condition on the components with respect to the observer's tetrad:
$$
  \frac{\d X^i}{\d \tau} = \frac{\d (X \cdot e^i)}{\d \tau} = 0
$$
implying that the four-vector $X$ varies according to:
$$
  \dot X = \sum_i (\dot{X^i e_i}) = \sum_i \dot X^i e_i + \sum_i X^i \dot e_i
  = \sum_i X^i \dot e_i = \sum_i X^i \Omega \cdot e_i = \Omega \cdot X
$$  
Similarly, for rank 2 tensors, one should have:
$$
  \frac{\d T^{ij}}{\d \tau} = 0
$$
implying:
$$
  \dot T = \sum_{ij} (\dot{T^{ij} e_i \otimes e_j}) = 
  \sum_{ij} T^{ij} \left( \dot e_i \otimes e_j + e_i \otimes \dot e_j \right)
  = \sum_{ij} T^{ij} \left( \Omega \cdot e_i \otimes e_j + e_i \otimes \Omega \cdot 
  e_j \right) = \Omega \cdot T + T \cdot \Omega^T
$$
where $^T$ stands for the transpose. Applying the above equation to the acceleration
tensor itself, one obtains that the condition of steadiness amounts to enforcing 
that the convective derivative of $\Omega$ at thermodynamical equilibrium must vanish,
i.e.:
\begin{equation}\label{omegadot}
      \dot \Omega = 0
\end{equation}

For a rigidly rotating fluid, all above conditions are fulfilled provided that the 
tetrad is the Frenet-Serret one. The local temperature $T_0$, the local chemical 
potential $\mu_0$ and the acceleration tensor have vanishing convective derivatives,
as it turns out from eqs.~(\ref{temp}),(\ref{mu}) and (\ref{omegatens}). The spin
tensor density is also stationary in the non-inertial frame because, for the ideal
gas, it is proportional to the acceleration tensor. We stress that this does not 
happen for other tetrads, e.g. a Fermi-Walker transported one; in that frame, the 
acceleration tensor components and, consequently, the spin density tensor components
would vary and therefore, one could not properly talk about local thermodynamical
equilibrium. This is why, in a rigidly rotating fluid, the Frenet-Serret frame
is a privileged reference.

Comparing the view of the inertial observer with that of the non-inertial one allows
to identify inertial forces and helps elucidating the role of spin. To start with, 
it is useful to take the non-relativistic limit of phase-space distribution (\ref{phsp2}) 
seen by the inertial observer. At first order in $v$, for a gas of spinless particles, 
this reads:
\begin{equation}\label{nonrel}
  f({\bf x},{\bf p}) \propto \exp \left[ - \frac{\p^2}{2mT} + 
  \frac{{\bf v}\cdot{\bf p}}{T} \right] = \exp \left[ - \frac{\p^2}{2mT} + 
  \frac{(\omegav \times {\bf x}) \cdot{\bf p}}{T} \right] = \exp \left[ - \frac{\p^2}{2mT} + 
  \frac{\omegav \cdot{\bf L}}{T} \right] 
\end{equation}
which, once integrated, yields the invariant partition function. At each point 
${\bf x}$ the inertial observer has to integrate in momentum the integral (\ref{nonrel}) 
because he deals with a rotating gas and, in order to take angular momentum 
conservation into account, he has to calculate a rotational grand-canonical partition 
function with an $\omegav \cdot {\bf L}$ term \cite{landau}. Conversely, for the 
non-inertial local observer, in the non-relativistic usual view, the particles of 
the gas as living in external fields, the centrifugal and Coriolis fields. He starts
writing a lagrangian:
\begin{equation}\label{lagra}
  L = \frac{1}{2} m {\bf v}'^2 + \frac{1}{2} m (\omegav' \times {\bf x})^2
  - m {\bf x} \cdot ({\omegav}' \times {\bf v}')
\end{equation}
with centrifugal and Coriolis potential energies. From (\ref{lagra}) a hamiltonian 
can be derived:
\begin{equation}\label{hami}
  H = \frac{1}{2} m {\bf v}'^2 - \frac{1}{2} m (\omegav' \times {\bf x})^2 =
  \frac{\p^2}{2m} - m {\bf p} \cdot (\omegav' \times {\bf x})
\end{equation}
where, from (\ref{lagra}):
\begin{equation}\label{varchan}
  {\bf p} = m {\bf v}' - m \omegav' \times {\bf x}
\end{equation}
Since in the non-relativistic limit $\omegav = \omegav'$, the non-inertial observer 
then writes a single-particle partition function in the point ${\bf x}$:
\begin{equation}\label{spfp}
 \int \d^3 \p \; \e^{-H/T} = \int \d^3 \p \; \exp \left[ - \frac{\p^2}{2mT} + 
  \frac{{\bf v}\cdot{\bf p}}{T} \right] 
\end{equation}
with the same temperature $T$, as there is no difference in the non-relativistic
limit. The result is the same the inertial observer would get, according to 
(\ref{nonrel}) and this means that pressure, entropy density and all local 
thermodynamical quantities are the same for the two observers. Yet, the physical 
interpretation of the non-inertial observer is different: for his point of view, 
the coupling between $\omegav$ and ${\bf x}\times{\bf p}$ arises from the presence 
of external fields, i.e. inertial forces, in his frame the angular momentum of 
the system is simply vanishing. 

It is worth pointing out that, from a relativistic point of view, the effect of 
centrifugal potential in the hamiltonian (\ref{hami}) is already contained in the 
dependence of temperature on the distance from the rotation axis. The rotating 
observers at different distances $r$ measure different local temperatures $T_0(r)$ 
and write the single-particle partition function of an ideal gas without external 
fields in their comoving frame, where $\beta=(1/T_0,{\bf 0})$:
\begin{equation}
 \int \d^3 \p \; \e^{-\beta \cdot p} = \int \d^3 \p \; \e^{-\varepsilon/T_0(r)} 
 = \int \d^3 \p \; \e^{-\varepsilon/\gamma (r) T} \simeq 
  m \int \d^3 v' \; \exp \left[-\frac{\varepsilon}{T} + \frac{\varepsilon v^2}{2T}
  \right] 
\end{equation}
The last term yields, in the non-relativistic limit and disregarding the mass term:
\begin{equation}
   m \int \d^3 {\rm v}' \; \exp \left[ -\frac{m v'^2}{2T} + \frac{1}{2} m 
  \frac{(\omegav \times {\bf x})^2}{T} \right]
\end{equation}
which eventually, after the change of variable (\ref{varchan}), turns out to
be equal to the right hand side of eq.~(\ref{spfp}). This result was as a consequence, 
from the point of view of the non-inertial observer, of the presence of an apparent 
gravitational field introducing a position-dependent temperature \cite{tolman}, 
which is still an inertial effect.

Similarly, the non-vanishing polarization of particles that he measures will be
attributed to an inertial effect, a coupling between the acceleration tensor 
and spin of the form:
$$
   - \frac{1}{2} \Omega_{\mu \nu} {\widehat S}^{\mu \nu} 
$$
which is, in his view, an additional term of the hamiltonian to be taken into account
when calculating thermodynamical quantities. Clarifying the nature of this energy, 
its relation with Thomas precession and spin-rotation coupling \cite{mashhoon2} and 
calculating thermodynamical quantities in stationary accelerated frames will
be the subjects of another work.

\section{Summary and outlook}

We have studied the physics of the ideal relativistic rotating gas at full 
thermodynamical equilibrium, in the Boltzmann approximation. We have shown that,
provided that it is sufficiently large, the gas is a rigidly rotating fluid with 
a non-vanishing spin density. We have obtained the expressions of the phase space
distribution (see Sect.~\ref{eqgas}):
$$
  f({\bf x},{\bf p})_{\tau \sigma} = \lambda \, \e^{-\beta \cdot p} 
  \, \frac{1}{2} \left( D^S([p]^{-1} {\sf R}_{\hat \omegav}(i \omega/T)[p])+
  D^S([p]^{\dagger} {\sf R}_{\hat \omegav}(i \omega/T) [p]^{\dagger-1}) 
  \right)_{\tau \sigma}
$$
as a matrix in the polarization states. From this distribution, we have inferred
the spin density tensor $\sigma_{\mu \nu}$ and found that, at equilibrium, is 
proportional to the acceleration tensor constructed with Frenet-Serret tetrad,
i.e.:
$$
            \sigma_{\mu \nu} = \iota \Omega_{\mu \nu}
$$
where $\iota$ is a Lorentz-scalar coefficient depending on the other thermodynamical
variables. Furthermore, we have recovered the generalized fundamental thermodynamical
relation
$$
 T_0 s = \rho - \mu_0 q + p - \frac{1}{2} \, \Omega_{\mu \nu} \sigma^{\mu \nu}	     
$$
including an additional term involving the spin density tensor, already used in
literature in fact. We have shown (see Appendix E) that this term cannot be made 
vanishing by using Belinfante transformations of the stress-energy and spin tensors
because this would imply a change in the total entropy and so the spin density
tensor has physical reality.
  
The condition of full equilibrium makes it possible to study a relativistic fluid
without the prerequirement of a theory of perfect fluids with spin and therefore, 
it allows to carry out an independent test of assumptions made in the past about
the dynamics of fluids with spin. Particularly, we have found that the projection
of the spin density tensor onto the four-velocity
$$
             t_\mu = \sigma_{\mu \nu} u^\nu
$$
is not vanishing, unlike commonly assumed \cite{vari,ec}, in agreement with general 
arguments of special relativity \cite{hagedorn} and envisaged by Bohm and Vigier
\cite{bohm}. Therefore, an observer comoving with the spin density needs two 
non-vanishing space-like vectors, not just one, to describe it.    

The non-vanishing of the four-vector $t$ has a considerable impact on the general
theory of a relativistic perfect fluid with spin, which needs to be revised from
the formulation of Halbwachs \cite{halb} and Ray \cite{ray}. This revision may 
open a window on new and unexpected features of relativistic hydrodynamics and 
its phenomenological consequences in various fields where relativistic hydrodynamics
is a fundamental tool, from relativistic heavy ion collisions to cosmology.

\section*{Acknowledgments}

We gratefully acknowledge useful discussions with G. Longhi, H. Liu, 
L. Lusanna, K. Rajagopal.
 


\appendix

\section*{APPENDIX A - Projection onto localized states}

The meaning of the projector $\Pro_V$ is easierly understood in a first-quantization 
framework for a single spinless particle, where it can be written a sum over a 
complete set of eigenfunctions vanishing outside the region $V$:
$$
    \Pro_V = \sum_{k} \ket{k}\bra{k}
$$
with:
$$
   u_{k}({\bf x}) = \braket{\bf x,\sigma}{k,\sigma} 
$$
The functions $u_k$ are complete in the sense that:
$$
  \sum_{k} u^*_{k}({\bf x}') u_{k}({\bf x}) = 
  \delta^3 ({\bf x}-{\bf x}') \theta_V({\bf x}) \theta_V({\bf x}')
$$
where $\theta_V({\bf x})=0$ if ${\bf x} \notin V$. Therefore, one has:
\begin{equation}
 \bra{{\bf p}'} \Pro_V \ket{{\bf p}} = \sum_k \braket{{\bf p}'}{k}
 \braket{k}{{\bf p}} = \int \d^3 \x \int \d^3 \x' \; 
 \e^{-i {\bf p}' \cdot {\bf x}'} \sum_{k} u_k ({\bf x}') u^*_k ({\bf x})
 \e^{i {\bf p} \cdot {\bf x} } =  \int_V \d^3 \x \; 
 \e^{i ({\bf p}-{\bf p}')\cdot {\bf x}} 
\end{equation}
which accounts for the integral part of eq.~(\ref{prov}).

In a quantum-field theoretical framework, the projector $\Pro_V$ can still be defined
and its matrix elements calculated \cite{beca1,beca2} giving rise to the 
eq.~(\ref{prov}) for single-particle states. Although the derivation of (\ref{prov})
in refs.~\cite{beca1,beca2} was partly based on an unproved conjecture, the 
non-trivial spin structure of eq.~(\ref{prov}) can be indeed justified by the 
requirement that the projector onto localized states commute with rotation operators:
$$
  [\Pro_V, \widehat{\sf R}_{\hat{\bf n}}(\phi)] = 0
$$
for any angle $\phi$. Therefore, by applying single-particle relativistic states on
both left and right hand side:
\begin{equation}\label{rhs}
  \bra{p',\sigma'} \Pro_V \widehat{\sf R}_{\hat{\bf n}}(\phi) \ket{p,\sigma}
  = \sum_{\tau} \bra{p',\sigma'} \Pro_V \ket{{\sf R}_{\hat{\bf n}} (\phi)(p),\tau} 
  D^S([{\sf R}_{\hat{\bf n}} (\phi)(p)]^{-1} {\sf R}_{\hat{\bf n}}(\phi) 
  [p])_{\tau \sigma}
\end{equation}
whereas:
\begin{equation}\label{lhs}
  \bra{p',\sigma'}\widehat{\sf R}_{\hat{\bf n}}(\phi) \Pro_V \ket{p,\sigma}
  = \bra{p,\sigma}\Pro_V^\dagger \widehat{\sf R}_{\hat{\bf n}}(\phi)^\dagger 
  \ket{p',\sigma'}^* = \sum_{\tau} \bra{p,\sigma} \Pro_V \ket{{\sf R}_{\hat{\bf n}}
  (\phi)^{-1}(p'),\tau}^* D^S([{\sf R}_{\hat{\bf n}}(\phi)^{-1}(p')]^{-1} 
  {\sf R}_{\hat{\bf n}}(\phi)^{-1} [p'])_{\tau \sigma'}^* 
\end{equation}
The (\ref{rhs}) and (\ref{lhs}) ought to be equal, so if we define:
$$
  M(p,q)_{\alpha,\beta} = \bra{p,\alpha} \Pro_V \ket{q,\beta}
$$
from (\ref{rhs}) and (\ref{lhs}) ensues:
\begin{equation}\label{condition}
 M(p',{\sf R}_{\hat{\bf n}}(\phi)(p)) 
 D^S([{\sf R}_{\hat{\bf n}} (\phi)(p)]^{-1} {\sf R}_{\hat{\bf n}}(\phi)[p])
 = D^S([p']^\dagger {\sf R}_{\hat{\bf n}}(\phi)[{\sf R}_{\hat{\bf n}}(\phi)
 ^{-1}(p')]^\dagger)
 M^\dagger(p,{\sf R}_{\hat{\bf n}}(\phi)^{-1}(p')) 
\end{equation}
The last equation tells us that the matrix $M$ cannot be proportional to the 
identity, as far as the spin is concerned. In fact, the eq.~(\ref{condition}) 
can be fulfilled if, according to eq.~(\ref{prov}) :
$$
  M(p,q) = F(p-q) \frac{1}{2} \left[ D^S([p]^{-1}[q]) + 
  D^S([p]^\dagger [q]^{\dagger-1}) \right]
$$
being $F$ a scalar function. Thereby, the eq.~(\ref{condition}) becomes:
\begin{eqnarray}
 && F(p'-{\sf R}_{\hat{\bf n}}(\phi)(p)) \frac{1}{2} \left[ D^S([p']^{-1}
  [{\sf R}_{\hat{\bf n}} (\phi)(p)]) + D^S([p']^\dagger [{\sf R}_{\hat{\bf n}}
  (\phi)(p)]^{\dagger-1}) \right] 
  D^S([{\sf R}_{\hat{\bf n}} (\phi)(p)]^{-1} {\sf R}_{\hat{\bf n}}(\phi)[p])
  \nonumber \\
 && = D^S([p']^\dagger {\sf R}_{\hat{\bf n}}(\phi)[{\sf R}_{\hat{\bf n}}(\phi)
 ^{-1}(p')]^{\dagger-1}) \frac{1}{2} \left[ D^S([{\sf R}_{\hat{\bf n}}
 (\phi)^{-1}(p')]^\dagger [p]^{\dagger -1}
  ) + D^S([{\sf R}_{\hat{\bf n}}(\phi)^{-1}(p)]^{-1}[p]) \right] 
  F(p-{\sf R}_{\hat{\bf n}}(\phi)^{-1}(p'))^* \nonumber \\
\end{eqnarray}
Taking advantage of the unitarity of Wigner rotations (see eq.~(\ref{unitary})), 
both left and right hand side of the above equation yield the same matrix 
multiplied by two different scalar functions:
\begin{eqnarray}
 && F(p'-{\sf R}_{\hat{\bf n}}(\phi)(p)) \frac{1}{2} \left( D^S([p']^{-1}
  {\sf R}_{\hat{\bf n}}(\phi)[p]) + D^S([p']^\dagger {\sf R}_{\hat{\bf n}}
  (\phi)[p]^{\dagger-1}) \right) \nonumber \\
 && = \frac{1}{2} \left( D^S([p']^{-1}
  {\sf R}_{\hat{\bf n}}(\phi)[p]) + D^S([p']^\dagger {\sf R}_{\hat{\bf n}}
  (\phi)[p]^{\dagger-1}) \right)
  F(p-{\sf R}_{\hat{\bf n}}(\phi)^{-1}(p'))^*
\end{eqnarray}
so that we are left with the simple requirement:
$$
 F(p'-{\sf R}_{\hat{\bf n}}(\phi)(p)) = F(p-{\sf R}_{\hat{\bf n}}(\phi)^{-1}(p'))^*
$$
which is fulfilled if, again according to eq.~(\ref{prov}):
$$
 F(p'-{\sf R}_{\hat{\bf n}}(\phi)(p)) \propto \int_V \d^3 \x \; \exp[i {\bf x}
 \cdot ({\bf p}'-{\sf R}_{\hat{\bf n}}(\phi)({\bf p}))]
$$
and if the region $V$ is invariant by rotation around the axis $\hat{\bf n}$.

\section*{APPENDIX B - Reality of phase space distribution}

We want to show that the integral:
\begin{equation}\label{realint}
 \int_V \d^3 \x \; \exp \left[ i {\bf x} \cdot ({\bf p}-
 {\sf R}_{\hat \omegav}(i \omega/T)^{-1}({\bf p})) \right] \;\; 
\end{equation}
is real. First, we note that:
$$
 {\bf p} - {\sf R}_{\hat \omegav}(i \omega/T)^{-1}({\bf p})
$$
has vanishing component along $\omegav$. Therefore, taking $\hat\omegav$ as the
unit vector of the $z$ axis, we can write the above integral as:
$$
 \int_V \d^3 \x \; \exp \left[ i {\bf x}_T \cdot ({\bf p}_T-
 {\sf R}_{\hat \omegav}(i \omega/T)^{-1}({\bf p}_T)) \right] \;\; 
$$
and, by using the explicit expression in (\ref{vectdiff}):
$$
 \int_V \d^3 \x \; \exp \left[ i {\bf x}_T \cdot \left( 
 (1-\cosh \omt){\bf p}_T - i \sinh \omt \, \hat \omegav \times {\bf p}_T
 \right) \right]
$$
The imaginary part of the last integral reads:
\begin{equation}\label{imint}
 \int_V \d^3 \x \; \sin \left[(1-\cosh \omt)\, {\bf x}_T \cdot {\bf p}_T
  \right] \exp\left[\sinh \omt \, {\bf x}_T \cdot (\hat \omegav \times {\bf p}_T)
  \right] 
\end{equation}
In order to show that this vanishes, we need to find a rotation of ${\bf x}_T$
around $\omegav$, i.e. in the transverse plane, which changes the sign of 
${\bf x}_T \cdot {\bf p}_T$ and, at the same time, leaves ${\bf x}_T \cdot 
(\hat \omegav \times {\bf p}_T)$ unchanged. This transformation in the 
transverse plane exists because ${\bf p}_T$ and $(\hat \omegav \times {\bf p}_T)$
belong to that plane and are orthogonal. Once this transformation is performed,
and the relevant change of variable ${\bf x} \to {\bf x}'$ in the integral
applied, the integral (\ref{imint}) is tranformed into its opposite as the
region $V$ is symmetric by rotation around $\omegav$, hence (\ref{imint})
vanishes.

\section*{APPENDIX C - The angular velocity vector in relativity}

To better understand the physical content of the antisymmetric tensor $\Omega$, 
it is useful to decompose it into two four-vectors through the definitions:
\begin{eqnarray}\label{content}
  A_\mu &=& \Omega_{\mu \nu} u^\nu \nonumber \\
 \omega_\mu &=& -\frac{1}{2} \epsilon_{\mu \nu \rho \sigma} \Omega^{\nu \rho} u^\sigma
\end{eqnarray}
which can be inverted to give \cite{misner}:
\begin{equation}\label{omega2}
 \Omega_{\mu \nu} = \epsilon_{\mu \nu \rho \sigma} \omega^\rho u^\sigma 
  + A_\mu u_\nu - A_\nu u_\mu
\end{equation}
Both four-vectors $A$ and $\omega$ are orthogonal to the four-velocity and are
therefore space-like. It is easy to verify, by using (\ref{omega1}) and reminding
that $e_0 = u$, that $A$ is just the four-acceleration vector $\d u/ \d \tau$.
In order to grasp the physical meaning of the four-vector $\omega$, it is necessary 
to first look at motion of the tetrad when $\omega=0$. In this case, the equation
of motion of the tetrad four-vectors read, from (\ref{tetrad4}) and (\ref{omega2}):
\begin{equation}\label{fw}
 \dot e_i^\mu  = A^\mu e_i \cdot u - u^\mu A \cdot e_i 
\end{equation}
This is the well known Fermi-Walker transport of the vector $e_i$. Now, with respect
to the Fermi-Walker transported tetrad, any other comoving spacial frame will be 
seen by that observer either as fixed with respect to the Fermi-Walker one or  
"rotating" with an equation like (\ref{classic}) with the time $t$ replaced by the 
observer's proper time $\tau$, that is:
\begin{equation}\label{fw2}
 \dot{\bf e}'_i = \omegav' \times {\bf e}'
\end{equation}
where the prime denotes that these are spacial vectors seen by the moving observer 
and not by the inertial one. In the Fermi-Walker transported frame the four-vector 
$\omega$ reads:
\begin{eqnarray}
 \omega_\mu &=& -\frac{1}{2} \epsilon_{\mu \nu \rho \sigma} \Omega^{\nu \rho} u^\sigma = 
 -\frac{1}{2} \epsilon_{\mu \nu \rho \sigma} \sum_{i=0}^3 \dot e^{i \nu} e^\rho_i u^\sigma 
 = -\frac{1}{2} \epsilon_{\mu \nu \rho \sigma} A^\nu u^\rho u^\sigma 
 -\frac{1}{2} \epsilon_{\mu \nu \rho \sigma} \sum_{i=1}^3 \dot e^{i \nu} e^\rho_i u^\sigma 
 \nonumber \\
 &=& -\frac{1}{2} \epsilon_{\mu \nu \rho 0} \sum_{i=1}^3 \dot e^{i \nu} e^\rho_i
\end{eqnarray}
being $u^\sigma = \delta^\sigma_0$ in any comoving frame. The last equality, along 
with (\ref{fw2}) implies that in the comoving Fermi-Walker transported frame 
$$
\omega = (0,\omegav')
$$
which makes it clear the meaning of the four-vector $\omega$. Now, to get it expressed 
in the inertial frame, it suffices to apply a pure Lorentz boost (without rotations) 
to the components seen by the comoving observer, with an instantaneuous velocity 
{\bf v} with respect to the inertial observer. Therefore:
\begin{eqnarray}\label{omegavect}
 \omega^0 &=& \gamma \omega'^0 + \gamma {\bf v} \cdot {\omegav}' =
 \gamma {\bf v} \cdot {\omegav}'  \nonumber \\ 
 \omega^i &=& \omega'^i + \gamma \omega'^0 v^i + \frac{\gamma -1}{v^2}
 v^j \omega'^j v^i = \omega'^i + \frac{\gamma -1}{v^2} v^j \omega'^j v^i
\end{eqnarray}
where we also used $\omega'^0=0$.

The last step is to provide a definition of angular velocity vector $\omegav$
in the inertial frame. We could, for instance, define it as the spacial part of the 
$\omega$ four-vector in (\ref{omegavect}), but it would not coincide with the 
angular velocity vector $\omegav$ that has been defined for a uniformly spinning 
system related to the velocity with eq.~(\ref{rigid}). Instead, we will use 
the latter as definition of $\omegav$ and thereafter derive its relation between
$\omegav'$. For a uniformly rotating system, we know the $\Omega$ tensor from
(\ref{omegatens}) and this allows to readily determine the $\omega$ four-vector
from (\ref{content}):
\begin{equation}\label{omegavect2}
  \omega = (\gamma^2 \omegav \cdot {\bf v}, \gamma^2 \omegav)
\end{equation}
We will keep this as a {\em definition} of $\omegav$ in general. One one hand, the 
(\ref{omegavect2}), along with the (\ref{timelike}), allows us to write the tensor 
$\Omega$ in general as a function of the acceleration ${\bf a}$ and the angular 
velocity $\omegav$ seen by the inertial frame, by using the (\ref{omega2}).
This leads to, in the inertial frame:
\begin{equation}\label{omegatensgen}
  \Omega = 
 \left( \begin{array}{c|c} 
    0 & \gamma^3 {\bf a} - \gamma^3 (\omegav \times {\bf v})  \\
    \hline
    - \gamma^3 {\bf a} - \gamma^3 (\omegav \times {\bf v}) &  
    \gamma^3 \omegav - \gamma^3 \omegav \cdot {\bf v} {\bf v}
    + \gamma^3 {\bf a} \times {\bf v}
  \end{array} \right)
\end{equation}
Finally, comparing (\ref{omegavect2}) with (\ref{omegavect}) we obtain the 
transformation rule of the angular velocity vector between the inertial and 
comoving frame:
\begin{equation}
 \omegav = \frac{1}{\gamma^2} \omegav' + 
 \frac{\gamma-1}{\gamma^2 v^2} {\bf v} \cdot {\omegav'} {\bf v} =
 \frac{1}{\gamma^2} \omegav' + 
 \frac{1}{\gamma + 1} {\bf v} \cdot {\omegav'} {\bf v}
\end{equation} 
%

\section*{APPENDIX D - Non-invariance of thermodynamics under Belinfante 
transformation}

The conservation of energy-momentum and angular momentum for a continuous 
system can be written as system of two partial differential equations:
\begin{eqnarray}
 && \partial_\mu T^{\mu \nu} = 0 \nonumber \\
 && \partial_\lambda {\cal J}^{\lambda, \mu \nu} = \partial_\lambda 
 ({\cal S}^{\lambda, \mu \nu} + x^\mu T^{\lambda \nu} - x^\nu T^{\lambda \mu})
 = 0 
\end{eqnarray}
$T$ being the stress-energy tensor and ${\cal S}$ the spin tensor. This 
set of two equations is invariant in form under a transformation \cite{halb}:
\begin{eqnarray}
 && T'^{\mu \nu} = T^{\mu \nu} +\frac{1}{2} \partial_\lambda
 \left( \Phi^{\lambda, \mu \nu } - \Phi^{\mu, \lambda \nu} - 
 \Phi^{\nu, \lambda \mu}  \right) \nonumber \\
 && {\cal S}'^{\lambda, \mu \nu} = {\cal S}^{\lambda, \mu \nu} -
 \Phi^{\lambda,\mu \nu}
\end{eqnarray}
where $\Phi$ is an arbitrary tensor of rank three which is antisymmetric in the
last two indices.

Choosing $\Phi^{\lambda, \mu \nu} = {\cal S}^{\lambda, \mu \nu}$, the new
spin tensor ${\cal S}'$ vanishes and the new stress-energy tensor
\begin{equation}\label{belinf}
 T'^{\mu \nu} = T^{\mu \nu} +\frac{1}{2} \partial_\lambda
 \left( {\cal S}^{\lambda, \mu \nu } - {\cal S}^{\mu, \lambda \nu} - 
 {\cal S}^{\nu, \lambda \mu}  \right)
\end{equation}
becomes symmetric; this is the well-known Belinfante symmetrization procedure. 
However, this transformation does not leave, for a system in full thermodynamical 
equilibrium, the basic relation (\ref{localfin}) invariant, whether the $\Omega:
\sigma/2$ term is included or not, and, moreover, it also changes the total 
entropy. To show this, we first remind the definitions of proper energy density 
and pressure for a given stress-energy tensor:
\begin{equation}\label{rhop}
  \rho = T^{\mu \nu} u_\mu u_\nu \qquad \qquad 
  p = \frac{1}{3} (u_\mu u_\nu - g_{\mu \nu}) T^{\mu \nu}
\end{equation} 
For a system at full thermodynamical equilibrium, such as a rigidly rotating fluid,
the stress-energy tensor {\em must} have \cite{israel} the form:
\begin{equation}\label{stressene}
  T^{\mu \nu} = (\rho + p) u^\mu u^\nu - p g^{\mu \nu}
\end{equation}
as well as a spin tensor which must factorize into a spin density second-order tensor
and the four-velocity:
\begin{equation}\label{spinten}
  {\cal S}^{\lambda, \mu \nu } = \sigma^{\mu \nu} u^\lambda
\end{equation} 
The stress-energy tensor in (\ref{stressene}) is already symmetric, yet the spin
tensor is not vanishing. We could then think of applying the Belinfante symmetrization
procedure described above in order to make the new spin tensor ${\cal S}'$ vanishing,
thereby obtaining a new symmetric stress-energy tensor $T'$. Using eq.~(\ref{belinf}) and 
(\ref{spinten}), this new tensor reads:
\begin{equation}\label{belinf2}
  T'^{\mu \nu} = T^{\mu \nu}+ \frac{1}{2} \partial_\lambda
 \left( \sigma^{\mu \nu } u^\lambda - \sigma^{\lambda \nu} u^\mu - 
 \sigma^{\lambda \mu} u^\nu \right) = T^{\mu \nu} - \frac{1}{2} 
 \partial_\lambda (\sigma^{\lambda \nu} u^\mu) - \frac{1}{2} 
 \partial_\lambda( \sigma^{\lambda \mu} u^\nu ) + \frac{1}{2} \dot 
 \sigma^{\mu \nu} + \frac{1}{2} \sigma^{\mu \nu} \partial \cdot u
\end{equation}
At thermodynamical equilibrium the last two terms vanish: the second one, because 
the velocity field is rigid, and the first one because of the (\ref{omegadot})
and the vanishing of convective derivatives of $T_0$, $\mu_0$:
$$
 \dot \sigma =\dot{(\iota \Omega)} = \dot \iota 
 (T_0,\mu_0,\Omega:\Omega) \, \Omega + \iota \dot \Omega = 0
$$  
Thus, the (\ref{belinf2}) becomes:
\begin{equation}\label{belinf3}
  T'^{\mu \nu} = T^{\mu \nu} - \frac{1}{2} 
 \partial_\lambda (\sigma^{\lambda \nu} u^\mu) - \frac{1}{2} 
 \partial_\lambda( \sigma^{\lambda \mu} u^\nu )
\end{equation}

For the stress-energy tensor according (\ref{belinf3}), the new energy density and 
pressure read, applying the (\ref{rhop}):
\begin{eqnarray}
 && \rho' = T'^{\mu \nu} u_\mu u_\nu = \rho - u_\mu u_\nu \partial_\lambda  
 (\sigma^{\lambda \nu} u^\mu) = \rho - u_\nu \partial_\lambda 
 \sigma^{\lambda \nu} \nonumber \\
 && p' = \frac{1}{3} (\rho' - T'^\mu_\mu) = \frac{1}{3} \left( \rho - u_\nu 
 \partial_\lambda \sigma^{\lambda \nu} - T^\mu_\mu + \partial_\lambda 
 (\sigma^{\lambda \nu} u_\nu) \right) = p + \frac{1}{3}
 \sigma^{\lambda \nu} \partial_\lambda u_\nu
\end{eqnarray}
The thermodynamical relation (\ref{localfin}) in the new variables should read, 
as ${\cal S}'=0$ and assuming the invariance of $q$:
\begin{equation}
 T_0 s' = \rho' + p' - \mu_0 q' = \rho + p -\mu_0 q - u_\nu \partial_\lambda 
\sigma^{\lambda \nu} + \frac{1}{3} \sigma^{\lambda \nu} \partial_\lambda u_\nu
\end{equation}
Hence, in order to keep the thermodynamic relation invariant, the entropy should
be modified by a term depending on the spin density tensor:
\begin{equation}
  T_0 s' = T_0 s - u_\nu \partial_\lambda 
\sigma^{\lambda \nu} + \frac{1}{3} \sigma^{\lambda \nu} \partial_\lambda u_\nu
 + \frac{1}{2} \Omega_{\lambda \nu} \sigma^{\lambda \nu} =
 T_0 s - \partial_\lambda (\sigma^{\lambda \nu} u_\nu) + \sigma^{\lambda \nu} 
 \partial_\lambda u_\nu + \frac{1}{3} \sigma^{\lambda \nu} \partial_\lambda u_\nu
 + \frac{1}{2} \Omega_{\lambda \nu} \sigma^{\lambda \nu}
\end{equation}
The last expression can also be rewritten as, by using the definition (\ref{tdef})
and taking advantage of the antisymmetry of $\sigma$:
\begin{equation}\label{deltaent}
  T_0 s' = T_0 s - \partial \cdot t + \frac{2}{3} \sigma^{\lambda \nu} (\partial_\lambda 
  u_\nu - \partial_\nu u_\lambda) + \frac{1}{2} \Omega_{\lambda \nu} 
  \sigma^{\lambda \nu}
\end{equation}
It can be shown that, for a rotating gas, all terms added to entropy density on the 
right-hand-side of the above equation are strictly positive. Using eqs.~(\ref{tvect}), 
(\ref{iotadef}) as well as (\ref{contrac}) and (\ref{omegatens}), the eq.~(\ref{deltaent}) 
can be rewritten as:
$$
  T_0 (s' - s ) = - \nabla \cdot (\iota \gamma^2 \omegav \times {\bf v}) + \frac{2}{3}
  \iota \gamma \omegav \cdot \nabla \times \gamma {\bf v} + \iota \gamma^2 \omega^2
$$
The vector field $\iota \gamma^2 \omegav \times {\bf v}$ is proportional to the 
centripetal acceleration and therefore has a negative divergence;
the curl of the vector field $\nabla \times \gamma {\bf v}$ for the velocity 
(\ref{rigid}) yields $ 2 \gamma \omegav$. Hence, all three terms on the right-hand are 
positive and total entropy should be changed by a finite positive amount.

The latter condition is clearly physically unacceptable. Since the rigidly rotating 
gas is a system in full thermodynamical equilibrium, the new entropy density obtained 
after the Belinfante transformation should at least have a vanishing integral  
over the system's region, which is not the case. 

Therefore, we conclude that making the spin tensor vanishing is impossible and it must 
have a physical reality.

\section*{APPENDIX E - The four-vector $t$}

The four-vector $t$, for a single massive particle, is defined as the expectation 
value of the operator $ \widehat \tau^\mu = \widehat J^{\mu \nu} \widehat P_\nu$ divided
by the mass. If the particle is in a pure state $\ket{p,\sigma}$ then the 
expectation value:
$$
 t^\mu \equiv \frac{1}{m} \bra{p,\sigma} \widehat \tau^\mu \ket{p,\sigma} = 
 \frac{1}{m} \bra{p,\sigma} \widehat J^{\mu \nu} \widehat P_\nu \ket{p,\sigma} 
$$
transforms as a vector because so does the relevant operator. In other words, 
by using the very definition of single particle states from the theory of 
Poincar\'e group representations \cite{moussa,weinberg}:
\begin{equation}\label{t1}
   t^\mu = \frac{1}{m} \bra{p^0,\sigma} \widehat {[p]}^{-1} 
   \widehat J^{\mu \nu} \widehat P_\nu \widehat{[p]} \ket{p^0,\sigma}
   = \frac{1}{m} \bra{p^0,\sigma} [p]^{\mu}_{\lambda}
   \widehat J^{\lambda \nu} \widehat P_\nu \ket{p^0,\sigma}
   = [p]^{\mu}_{\lambda} \frac{1}{m} \bra{p^0,\sigma} 
   \widehat J^{\lambda \nu} \widehat P_\nu \ket{p^0,\sigma}
\end{equation}
where $p^0 = (m,0,0,0)$.
Therefore, to prove that $t=0$ we just need to show it for a particle at rest.
In this case:
$$
 \bra{p^0,\sigma} \widehat J^{\mu \nu} \widehat P_\nu \ket{p^0,\sigma}
 = m \bra{p^0,\sigma} \widehat J^{\mu 0} \ket{p^0,\sigma}  
$$
The boost generators $\widehat J^{\mu 0}$ however anticommute with the parity 
operator and since the states of a single particle at rest have a definite 
parity (which is, by definition, the intrinsic parity of particles) the right
hand side of the above equation ought to vanish. Thence, in view of (\ref{t1}):
\begin{equation}\label{t2}
  t^\mu = \frac{1}{m} \bra{p,\sigma} \widehat \tau^{\mu} \ket{p,\sigma} = 
  \frac{1}{m} \bra{p,\sigma} \widehat J^{\mu \nu} \widehat P_\nu
  \ket{p,\sigma} = 0
\end{equation}

The validity of eq.~(\ref{t2}) is readily extended to {\em any pure state}. 
First we note that the above reasoning can be extended to a matrix element
with different spin components and same momentum:
$$
 \bra{p,\sigma} \widehat \tau^\mu \ket{p,\sigma'} = 0
$$
For a general pure state, it suffices to decompose single particle states onto 
momentum-spin eigenstates and show that:
\begin{equation}\label{t3}
 \bra{p',\sigma'} \widehat \tau^\mu \ket{p,\sigma} = 0
\end{equation}
for $p \ne p'$. By using standard Poincar\'e group commutation relations:
$$
  [\widehat J^{\lambda \sigma}, \widehat P^{\nu}] = i \widehat P^\lambda
  g^{\nu \sigma} -i \widehat P^\sigma g^{\nu \lambda}
$$  
it is easy to show that:
\begin{equation}
      [\widehat \tau^\mu, \widehat P^\nu] = i \widehat P^\mu \widehat P^\nu
      - i g^{\mu \nu} \widehat P^2
\end{equation}
and therefore, if $p \ne p'$:
$$
 \bra{p',\sigma'} \widehat \tau^\mu \widehat P^\nu \ket{p,\sigma} =
 \bra{p',\sigma'} \widehat P^\nu \widehat \tau^\mu \ket{p,\sigma} =
 p^\nu \bra{p',\sigma'} \widehat t^\mu \ket{p,\sigma}
= p'^\nu \bra{p',\sigma'} \widehat t^\mu \ket{p,\sigma}
$$
that is:
$$
 (p^\nu - p'^\nu) \bra{p',\sigma'} \widehat \tau^\mu \ket{p,\sigma} = 0
$$
Since the last equation applies to all $p$'s components, and $p \ne p'$, it
turns out that the only possibility is that the matrix element of $\widehat
t$ vanishes, i.e. the eq.~(\ref{t3}).

Altogether, all single-particle matrix elements of the operator $\widehat \tau$
vanish, i.e.:
$$
 \bra{p',\sigma'} \widehat \tau^\mu \ket{p,\sigma} = 0
$$
for any $p$, $p'$, $\sigma$ and $\sigma'$, and thus the expectation value
of $\widehat \tau$, hence $t$, vanishes for a generic pure state $\ket{\Phi}$:
$$
   t^{\mu} = \frac{1}{m} \bra{\Phi} \widehat \tau^\mu \ket{\Phi} = 0
$$

However, for a free multiparticle system this is no longer true \cite{hagedorn} 
because:
$$
  \widehat \tau^\mu = \widehat J^{\mu \nu} \widehat P_\nu =
  \sum_i \widehat J^{\mu \nu}_i \sum_j \widehat P_{j \nu} \ne
  \sum_i \widehat t_i^\mu
$$
Also, for a single particle in a mixture of states the expectation value of
$\widehat \tau$, i.e. $\tr (\widehat \rho \widehat \tau)$ is not vanishing unless 
the density operator is a scalar operator, i.e. commutes with all unitary 
representations of Lorentz transformations.
In fact, the operator $\exp[\omegav \cdot \widehat{\bf J}/T]$, that we have
dealt with in this work, does not.

\section*{APPENDIX F - Polarization four-vector}

The results of Sect.~\ref{spin} allow to determine the polarization four-vector of 
particles with momentum ${\bf p}$ and general spin $S$, in an ideal Boltzmann gas. 
This determination is alternative with respect to that presented in 
ref.~\cite{becapicc}. The polarization four-vector is defined as:
\begin{equation}\label{pola1}
 \Pi^\mu (p) \equiv \frac{1}{m} \sum_\sigma \bra{p,\sigma} 
 \widehat W^\mu \widehat \rho \ket{p,\sigma} \equiv
 \frac{1}{m} \langle\langle \widehat W^\mu \rangle\rangle_p
\end{equation}
where $W$ is the Pauli-Lubanski four-vector:
$$
 W_\mu = -\frac{1}{2} \epsilon_{\mu \nu \rho \tau} J^{\nu \rho} P^\tau
$$
and $\widehat \rho$ is the single-particle spin density operator, which, for our 
present case reads:
$$
\widehat \rho = \frac{1}{n({\bf p})} 
\exp[-(\varepsilon({\bf p}) + \mu \widehat q + \omegav \cdot \widehat 
{\bf j})/T] \Pro_V
$$
being 
$$
 n({\bf p}) = \sum_\sigma \bra{p,\sigma} 
 \exp[-(\varepsilon({\bf p}) + \mu \widehat q + \omegav \cdot \widehat {\bf j})/T] 
 \Pro_V \ket{p,\sigma} = \int_V \d^3 \x \; f({\bf x},{\bf p})
$$
Thus, in view of (\ref{pola1}), the polarization four-vector can be written in the
inertial frame as:
\begin{eqnarray}\label{pola2}
 \Pi_0 &=& \frac{1}{n({\bf p})} \langle\langle \widehat{\bf j} \rangle \rangle_p
 \cdot \frac{{\bf p}}{m}  \nonumber \\
 \Piv &=& \frac{1}{n({\bf p})} \langle\langle \widehat{\bf k} \rangle \rangle_p 
 \times \frac{{\bf p}}{m} + \frac{1}{n({\bf p})} \langle\langle \widehat{\bf j} 
 \rangle \rangle_p \frac{\varepsilon}{m} 
\end{eqnarray}
By using the results of eqs.~(\ref{rota3}),(\ref{sdens1}) and (\ref{orboost4}), 
we can readily find the various expression of the expectation values in the 
eq.~(\ref{pola2}), that is:
\begin{eqnarray}\label{expval}
 \langle\langle \widehat{\bf j} \rangle \rangle_p &=&
 \int_V \d^3 \x \; ({\bf x} \times {\bf p}) f({\bf x},{\bf p}) + 
 \int_V \d^3 \x \; f({\bf x},{\bf p}) \frac{\chi'\! \left( \omt \right)}{\chiomt} 
 \hat\omegav \nonumber \\
 \langle\langle \widehat{\bf k} \rangle \rangle_p &=&
 \int_V \d^3 \x \; (C {\bf p} - \varepsilon {\bf x}) f({\bf x},{\bf p})
\end{eqnarray}
By plugging (\ref{expval}) into (\ref{pola2}) we get:
\begin{eqnarray}
 \Pi_0 &=& \frac{\chi'\! \left( \omt \right)}{\chiomt} \hat\omegav \cdot {\bf p} 
 \nonumber \\
 \Piv &=& \frac{\chi'\! \left( \omt \right)}{\chiomt} \frac{\varepsilon}{m}
 \hat\omegav 
\end{eqnarray}
which is the same result found in ref.~\cite{becapicc}.

\end{document}